\documentclass[a4paper,11pt,dvipsnames]{article}
\pdfoutput=1 %
\usepackage{jheppub} %
\RequirePackage{graphicx}

\usepackage{amsfonts,amsmath,amssymb}
\usepackage[utf8]{inputenc}
\usepackage{comment}
\usepackage{caption}
\usepackage{subcaption}
\usepackage{braket}

\DeclareMathOperator\arctanh{arctanh}
\DeclareMathOperator\sgn{sgn}
\newcommand{\gmt}{$g$$-$$2$ }
\newcommand{\had}{\mathrm{had}}
\newcommand{\lep}{\mathrm{lep}}
\newcommand{\muone}{\textsc{MUonE} }
\newcommand{\D}{\mathcal{D}}
\newcommand\uprule{\rule{0mm}{2.0ex}}

\begin{document}
\preprint{\vbox{\hbox{SI-HEP-2018-28 \quad QEFT-2018-17}}}

\title{\boldmath Hadronic corrections to $\mu$-$e$ scattering\\ at NNLO with space-like data}

\author{Matteo Fael}
\emailAdd{fael@physik.uni-siegen.de}
\affiliation{Theoretische Physik I, Universit\"at Siegen, Walter-Flex-Strasse 3, 57068 Siegen, Germany}

\abstract{
The Standard Model prediction for $\mu$-$e$ scattering at Next-to-Next-to-Leading Order (NNLO) contains non-perturbative QCD contributions given by diagrams with a hadronic vacuum polarization insertion in the photon propagator. 
By taking advantage of the hyperspherical integration method, we show that the subset of hadronic  NNLO corrections where the vacuum polarization appears inside a loop, the irreducible diagrams, can be calculated employing the hadronic vacuum polarization in the space-like region, without making use of the $R$ ratio and time-like data. 
We present the analytic expressions of the kernels necessary to evaluate  numerically the two types of irreducible diagrams: the two-loop vertex and box corrections.
As a cross check, we evaluate these corrections numerically and we compare them with the results given by the traditional dispersive approach and with analytic two-loop vertex results in QED.
}

\maketitle
\flushbottom

\section{Introduction}
\label{sec:intro}

The goal of the new \gmt experiments at Fermilab in the United States and at J-PARC in Japan is to measure the muon anomalous magnetic moment with a precision of $1.6 \times 10^{-10}$~\cite{Grange:2015fou,Saito:2012zz} --- corresponding to 140 ppb --- an improvement by a factor of four of the final BNL E821 experiment's uncertainty: $\delta a_\mu^\mathrm{exp} = 6.3 \times 10^{-10}$ (540 ppb) \cite{Bennett:2006fi}.
By all means, the theoretical prediction must keep up with the experimental precision. 
At present, the Standard Model prediction of the muon \gmt is limited by the uncertainty of the Hadronic Leading Order (HLO) and Light-By-Light (HLBL) contributions, that cannot be computed in perturbative QCD. 

The most precise determinations of $a_\mu^\mathrm{HLO}$, the leading hadronic contribution to the muon \gmt, are calculated employing the very well-known dispersive integral~\cite{Bouchiat1961,Durand:1962zzb,Gourdin:1969dm}
\begin{equation}
  a_\mu^\mathrm{HLO} = 
  \left( \frac{\alpha m_\mu}{3 \pi} \right)^2
  \int_{4m_\pi^2}^\infty
  \frac{ds}{s^2} R(s) \hat K(s),
  \label{eqn:disprel}
\end{equation}
and the ratio $R(s) = \sigma(e^+e^- \to \gamma^* \to \mathrm{hadrons})/\frac{4\pi\alpha^2}{3s}$ that can be measured at low energies. 
The kernel $\hat K(s)$ is a monotonically increasing function, with $\hat K(4m_\pi^2) \simeq 0.63$ increasing to one at $s \to +\infty$.
The present error on $a_\mu^\mathrm{HLO}$ is about $3 \times 10^{-10}$~\cite{Jegerlehner:2017zsb,Davier:2017zfy,Keshavarzi:2018mgv} --- a relative accuracy of $0.6\%$ --- and constitutes roughly 50\% of the Standard Model (SM) error budget.
Even if so far with a larger uncertainty, lattice QCD provides an alternative evaluation of $a_\mu^\mathrm{HLO}$~\cite{Aubin:2006xv,Boyle:2011hu,Feng:2011zk,DellaMorte:2011aa,Blum:2015you,Chakraborty:2016mwy,Borsanyi:2016lpl,Borsanyi:2017zdw,Meyer:2018til}. 
Other proposed methods exploit the hadronic vacuum polarization within dispersive QCD approach~\cite{Nesterenko:2014txa}, Schwinger's sum rules~\cite{Hagelstein:2017obr} and Mellin-Barnes approximants~\cite{Charles:2017snx}.

Recently a new experiment, \muone, has been proposed at CERN to determine $a_\mu^\mathrm{HLO}$ by measuring the running fine-structure constant $\alpha$,
\begin{equation}
  \alpha(q^2) = \frac{\alpha}{1+\mathrm{Re} \, \Pi(q^2)},
\end{equation}
in the space-like region ($q^2<0$) in $\mu$-$e$ scattering as a function of the squared momentum transfer $t$~\cite{Calame:2015fva,Abbiendi:2016xup}.
The function $\Pi(q^2)$ is the renormalized photon vacuum polarization from which it is possible to extract the hadronic contribution $\Pi^\had(q^2)$ by subtracting from $\Pi (q^2)$ the leptonic part $\Pi^\lep(q^2)$, calculated in perturbative QED up to four loops~\cite{Meyer:2018til}.\footnote{The sharp separation between $\Pi^\lep(q^2)$ and $\Pi^\had(q^2)$ is valid only up to two loops. Starting from three loops there are diagrams with both leptons and hadrons. Note that the vacuum polarization $\Pi (q^2)$ receives contributions also from the top quark and $W$ boson.}
The hadronic vacuum polarization $\Pi^\had(q^2)$ in the space-like region can provide an independent determination of $a_\mu^\mathrm{HLO}$ thanks to the formula~\cite{Lautrup:1971jf}
\begin{equation}
a_\mu^\mathrm{HLO} = 
-\frac{\alpha}{\pi}
\int_0^1 dx \, (1-x) \, \Pi^\had
\left( \frac{m_\mu^2 x^2}{x-1} \right),
\label{eqn:aHLOspace}
\end{equation}
To determine $a_\mu^\mathrm{HLO}$ with an error of about $2 \times 10^{-10}$ the \muone experiment must measure the differential cross section with statistic and systematic uncertainties of the order of 10 ppm.

To this end, a Monte Carlo event generator that includes all relevant corrections needed to reach such level of precision must be developed.
It must contain QED and QCD  radiative  corrections up to Next-to-Next-to-Leading-Order (NNLO) in $\alpha$ matched to leading-logarithmic corrections resummed to all orders.
A Next-to-Leading-Order (NLO) Monte Carlo generator based on the existing \textsc{BabaYaga}~\cite{CarloniCalame:2000pz,CarloniCalame:2001ny,CarloniCalame:2003yt,Balossini:2006wc,Balossini:2008xr} framework is currently under development~\cite{Alacevich:2018vez}.
A first step towards the evaluation of NNLO corrections was presented in~\cite{Mastrolia:2017pfy,DiVita:2018nnh} where the QED two-loop master integrals were calculated for finite muon mass and vanishing electron mass. 

\begin{figure}[htb]
  \centering
  \begin{minipage}[b]{0.24\textwidth}
    \includegraphics[width=\textwidth]{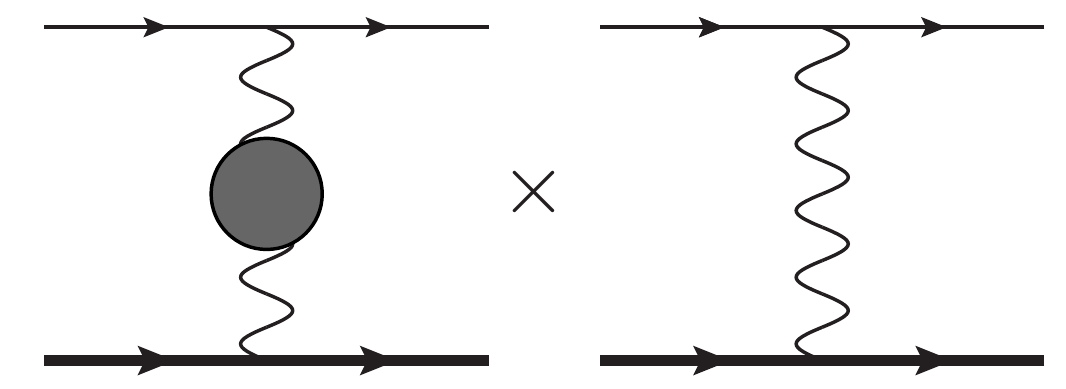}

    \vspace*{0.3\textwidth}
    \subcaption{Hadronic\\ contribution at NLO}
    \label{fig:hadLO}
  \end{minipage}\quad
  \begin{minipage}[b]{0.24\textwidth}
  \begin{tabular}{c}
   \includegraphics[width=\textwidth]{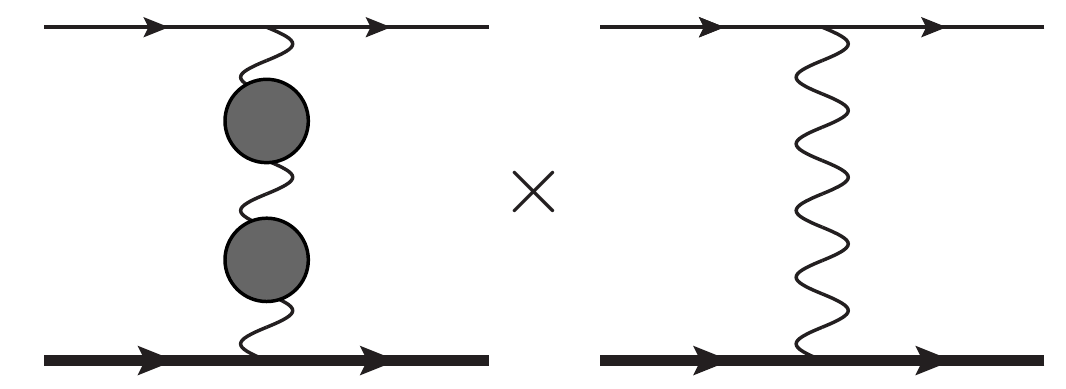}\\
   \includegraphics[width=\textwidth]{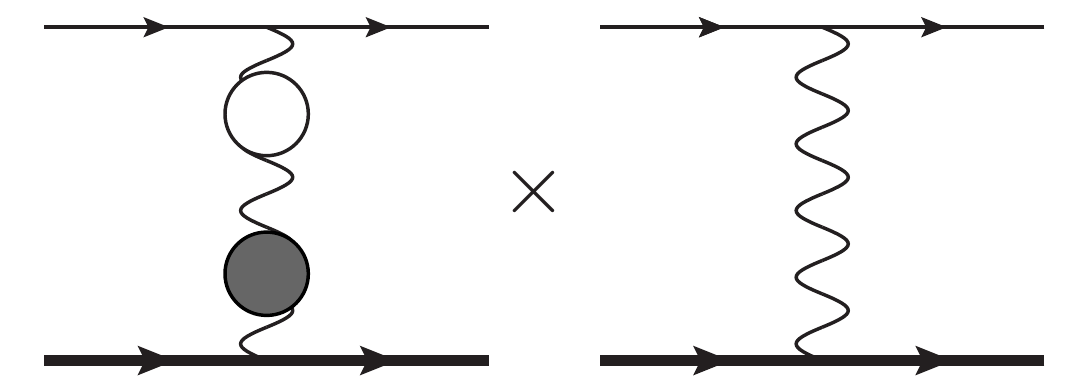}\\
   \includegraphics[width=\textwidth]{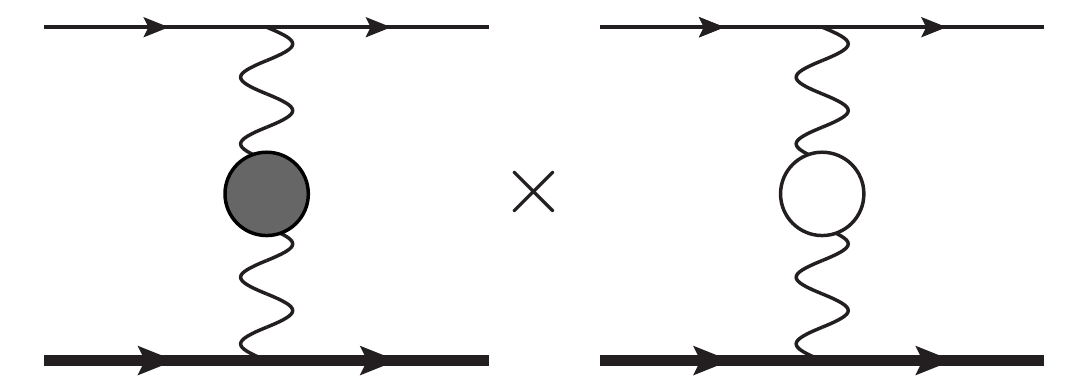} 
  \end{tabular} 
    \subcaption{Class I\\ }
    \label{fig:classI}
  \end{minipage}\quad
  \begin{minipage}[b]{0.4\textwidth}
  \begin{tabular}{c}
    \includegraphics[width=\textwidth]{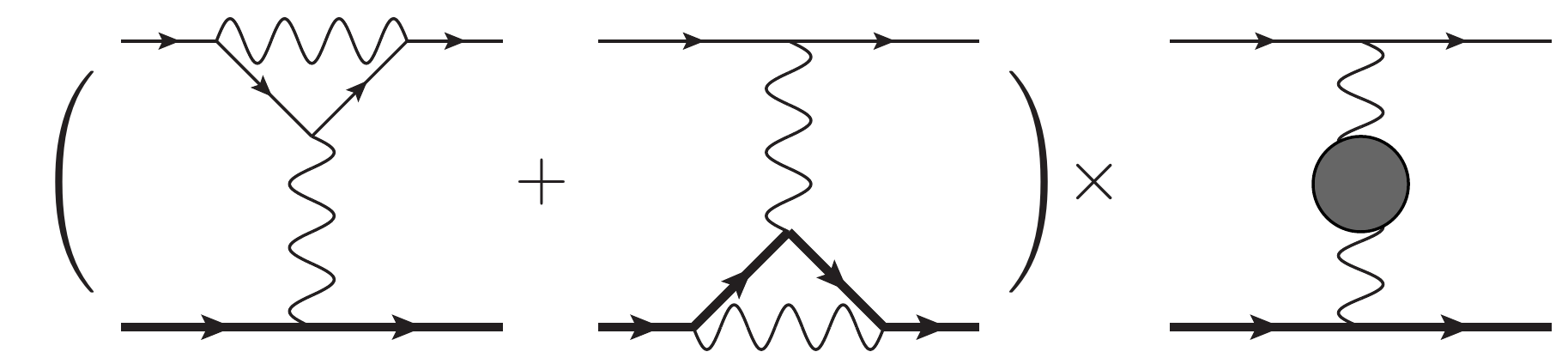}\\
    \includegraphics[width=\textwidth]{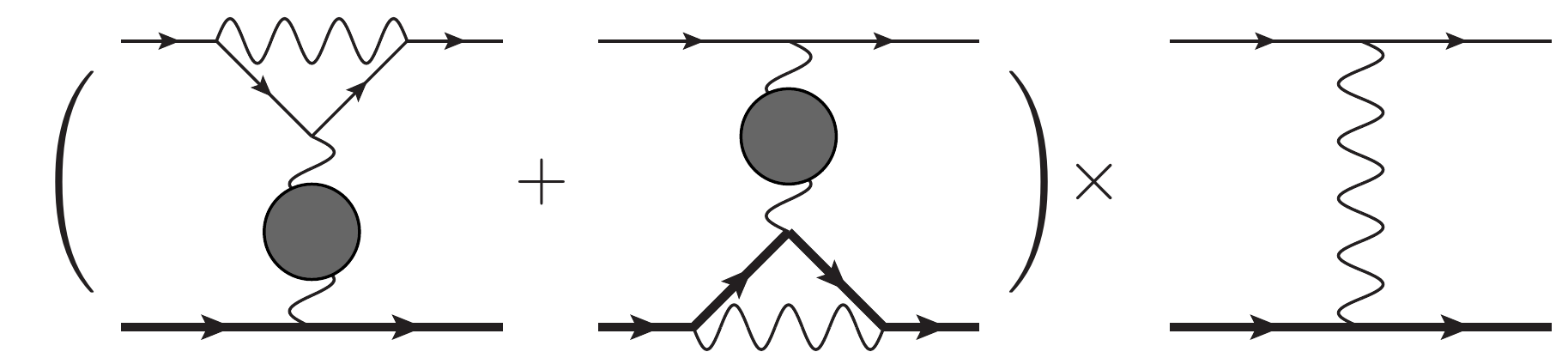}\\
    \includegraphics[width=\textwidth]{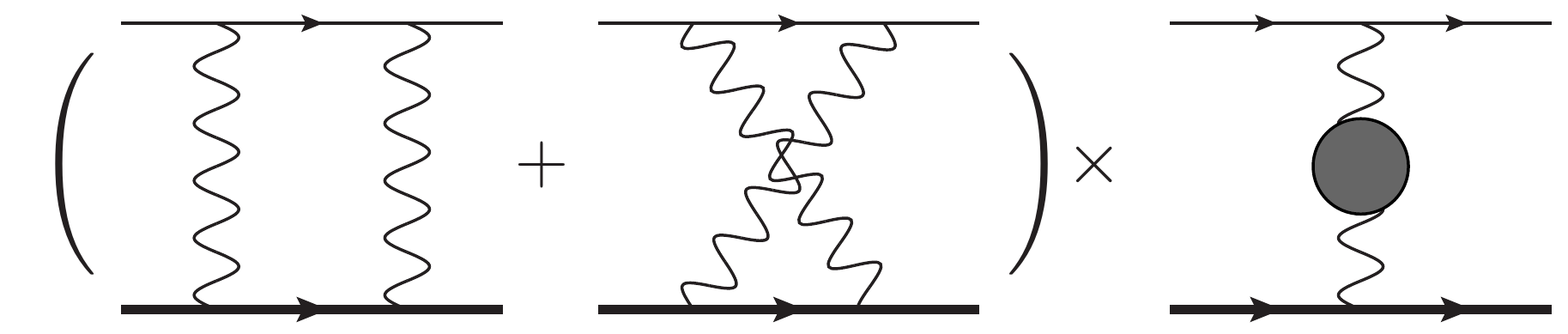}
  \end{tabular} 
    \subcaption{Class II\\ }
    \label{fig:classII}
  \end{minipage} \\[10pt]
  \begin{minipage}[b]{0.48\textwidth}
    \centering
    \includegraphics[width=\textwidth]{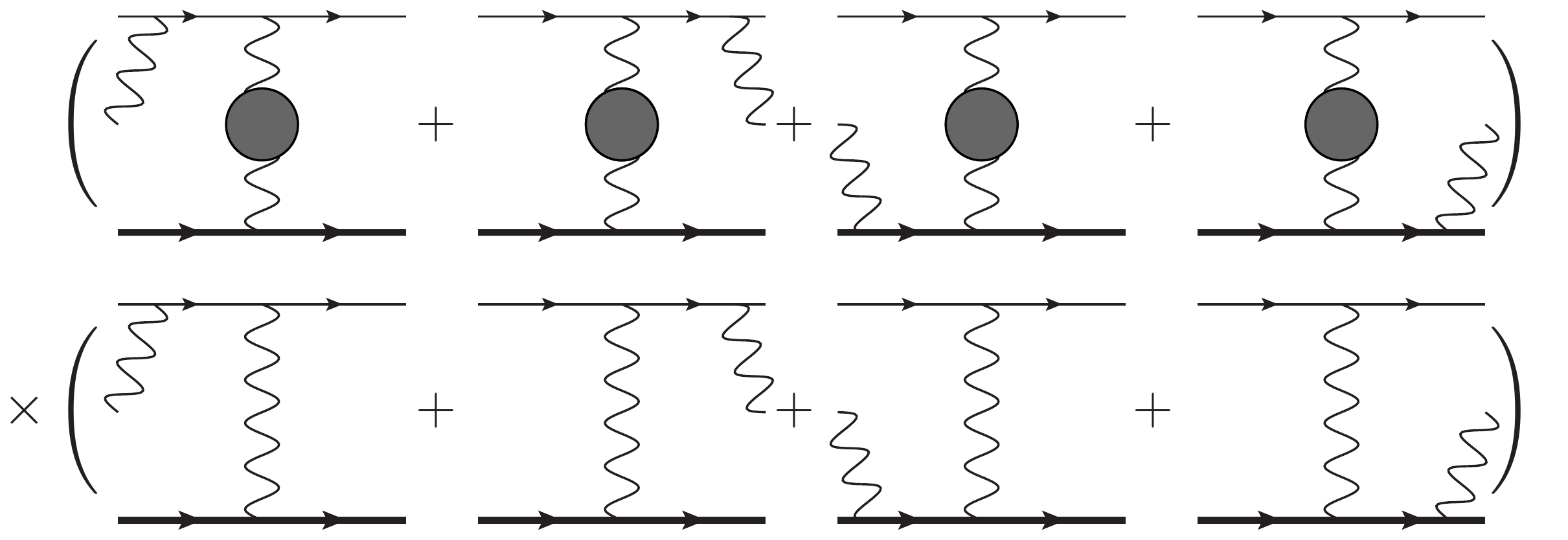}

    \vspace*{0.15\textwidth}
    \subcaption{Class III}
    \label{fig:classIII}
  \end{minipage}\quad\quad
  \begin{minipage}[b]{0.4\textwidth}
      \begin{tabular}{c}
      \includegraphics[width=\textwidth]{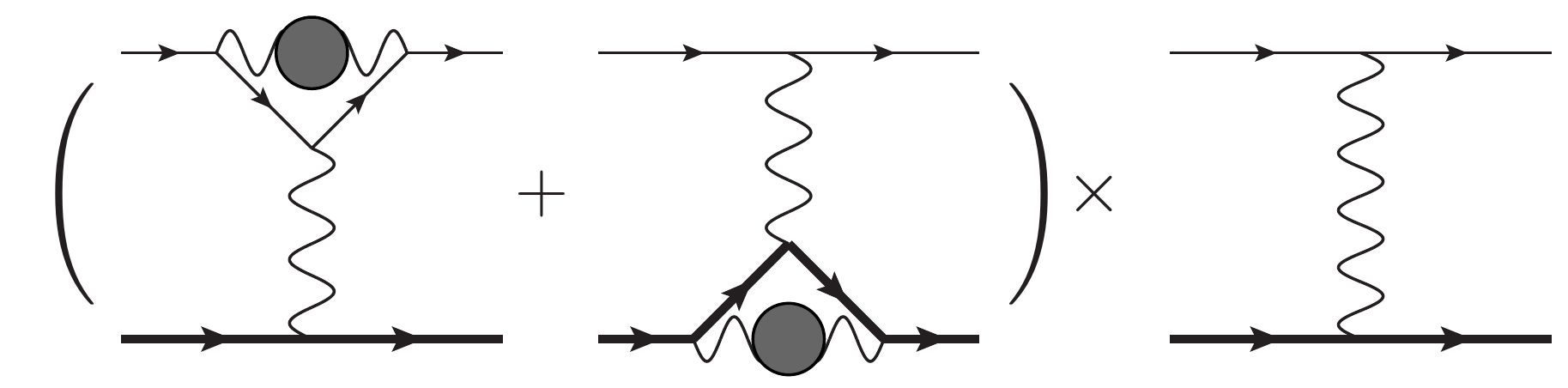} \\
      \includegraphics[width=\textwidth]{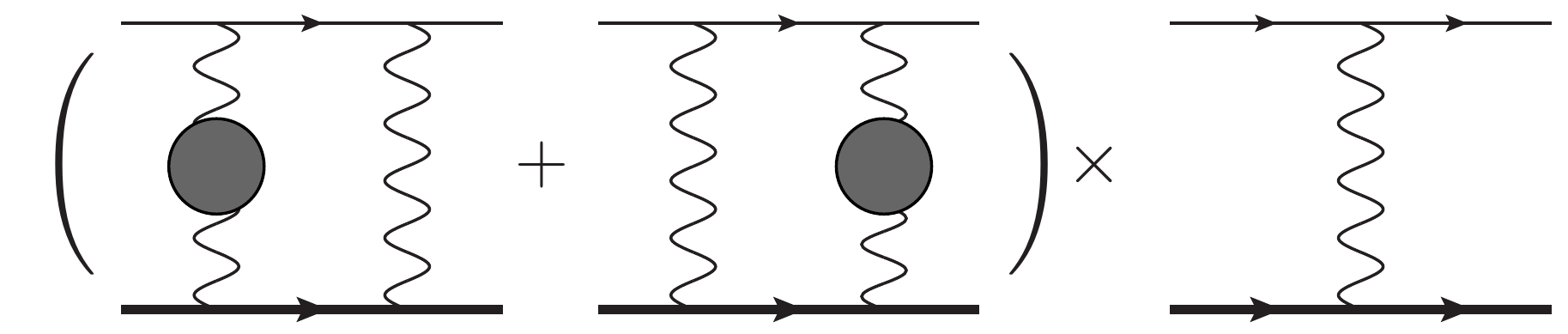} \\
      \includegraphics[width=\textwidth]{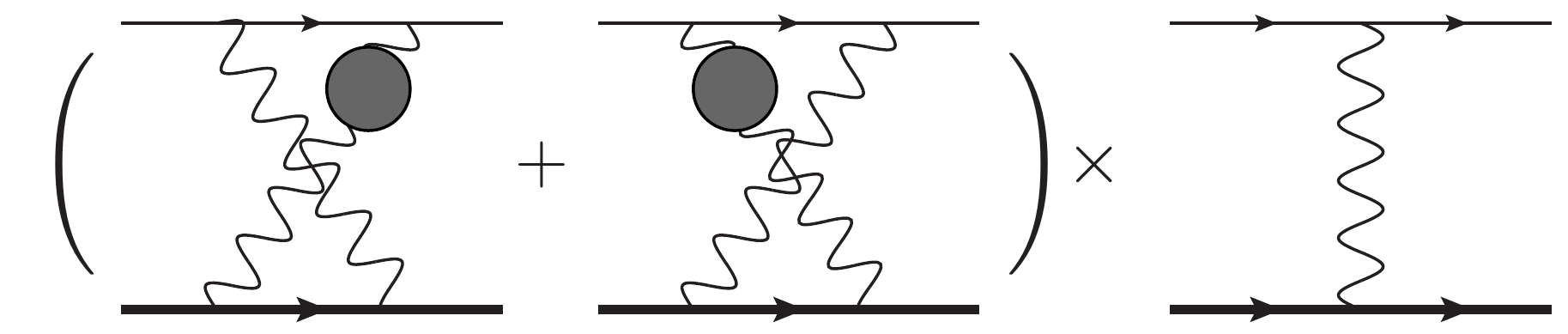}
      \end{tabular}
    \subcaption{Class IV}
    \label{fig:classIV}
  \end{minipage}
  \caption{The hadronic contributions to $\mu$-$e$ scattering at NLO (a) and at NNLO (b-e). 
  Muon and electron lines are drawn with thick and thin lines, respectively.}
  \label{fig:haddiagrams}
\end{figure}

In this paper we will focus on the hadronic contributions to the $\mu$-$e$ scattering cross section. These corrections are genuinely non-perturbative and cannot be calculated in perturbative QCD since the scattering process will take place at a center-of-mass energy of about $0.5$~GeV.
The hadronic contribution to the NLO cross section --- order $\alpha^3$ --- comes from the diagram in figure~\ref{fig:hadLO}; 
it corresponds to the leading effect of the fine-structure-constant running in an expansion of $\alpha=\alpha(0)$.

The hadronic corrections to the NNLO cross sections --- order $\alpha^4$ --- can be divided into four classes of diagrams.
 \begin{enumerate}
   \item[I.] Tree-level diagrams with double vacuum polarization insertion (figure~\ref{fig:classI}), either two hadronic insertions or one $\Pi^\had$ and one $\Pi^\lep$. These are the second order effects of the running of $\alpha$.
   \item[II.] QED one-loop diagrams in combination with one insertion of $\Pi^\had$ in the $t$-channel photon (figure~\ref{fig:classII}). 
Their contribution to the cross section is linear in $\Pi^\had(t)$ and can be obtained directly from the QED one-loop amplitudes.
  \item[III.] Real photon emission with a dressed photon propagator in the $t$ channel (figure~\ref{fig:classIII}). 
\end{enumerate}
All the diagrams in class I-III are \emph{factorizable} or \emph{reducible} since they are given by the product of a QED amplitude times the function $\Pi^\had(q^2)$ evaluated at $q^2=t$. A fourth class of \emph{non-factorizable} or \emph{irreducible} diagrams must be also considered:
\begin{enumerate}
  \item[IV.] One-loop QED amplitudes with a dressed photon propagator inserted inside the loop. They can be further subdivided into vertex and box corrections (figure~\ref{fig:classIV}).
\end{enumerate}
Note that there is no LBL contribution to the cross section up to N$^3$LO --- order $\alpha^5$.\footnote{In figure~\ref{fig:hadLO} a virtual photon can be emitted and reabsorbed by the hadronic bubble. In the spirit of the common nomenclature, we do not consider this kind of two-loop diagrams as a part of the hadronic NNLO corrections.
This effect is commonly included in $R(s)$ as final state radiation, so no additional contribution has to be taken into account.} 
Moreover, we remind the reader that the analysis of future \textsc{MUonE} data will also require the study of $\mu$-$e$ scattering processes with final states containing hadrons. Final states of Bhabha scattering containing hadrons were studied in~\cite{CarloniCalame:2011zq}.

The traditional approach to calculate the amplitudes in class IV uses the dispersion relation,
\begin{equation}
  \frac{\Pi^\had(q^2)}{q^2} =
  -\frac{\alpha}{3\pi}
  \int_{4m_\pi^2}^{\infty}
  \frac{dz}{z} \frac{R(z +i \varepsilon)}{q^2-z+i\varepsilon} \, ,
  \label{eqn:dispersionrelationR}
\end{equation}
to replace the dressed photon propagator inside the loop --- now $q$ stands for the loop momentum --- with the r.h.s.\ of eq.~\eqref{eqn:dispersionrelationR} where the momentum $q$ appears only in the term $1/(q^2-z)$.
This allows us to interchange the integration order and calculate as a first step the one-loop integrals with the dressed photon propagator replaced by a massive gauge boson of mass $\sqrt{z}$.
Later on the $z$-dependent scattering amplitudes are convoluted with the $R$~ratio.
The dispersive approach was employed for instance to calculate the hadronic corrections to Bhabha scattering~\cite{Actis:2007fs,Kuhn:2008zs}.
A complete calculation of the hadronic corrections to $\mu$-$e$ scattering at NNLO with the dispersive approach will be presented soon~\cite{Fael:2019nsf}.

The dispersive approach requires the $R$~ratio as an input. 
Therefore the \textsc{MUonE}'s determination of $\Pi^\had(q^2)$  and $a_\mu^\mathrm{HLO}$ from space-like data would still depend marginally on time-like data if the dispersive approach were employed in the evaluation of the hadronic NNLO corrections.
The alternative  is to use the very same space-like data measured by \muone to calculate the hadronic NNLO corrections iteratively without dispersion relation and the $R$~ratio. 
One could approximate the function $\Pi^\had(t)$ through successive iterations: as a first step the hadronic NNLO corrections can be switched off in the Monte Carlo  and a first approximation for $\Pi^\had(t)$ extracted.
Afterwards, the Monte Carlo can be supplied with this first approximation to evaluate the hadronic NNLO corrections, a second approximation calculated and the process further iterated.
The factorizable diagrams in class I, II and III depend on $\Pi^\had(t)$, so they are well suited to implement this iterative procedure. 
What about the contributions of the irreducible diagrams in class IV?

In this paper we will show that also the non-factorizable diagrams can be calculated using the hadronic vacuum polarization in the space-like region by making use of the hyperspherical integration method.
This method was exploited for instance to evaluate parts of the QED three-loop contributions to the electron \gmt~\cite{Levine:1973un,Levine:1974xh,Levine:1979uz,Roskies:1990ki,Laporta:1991zw,Laporta:1992pa}, and more recently to calculate the pion pole contribution to $a_\mu^\mathrm{HLBL}$\cite{Knecht:2001qf,Jegerlehner:2009ry} and in the dispersive approach to the HLBL~\cite{Colangelo:2014dfa,Colangelo:2014pva}.

The loop integrals containing $\Pi^\had(q^2)$ can be calculated as follows: after analytic continuation of internal and external momenta to the Euclidean region, one introduces spherical coordinates for the loop momentum $q$.
The angular dependence of the Feynman propagators can be made explicit by an expansion in Gegenbauer polynomials.
Afterwards, the integration with respect to the angular variables can be carried out analytically by taking advantage of the orthogonality properties of these polynomials. 
In this way, the non-factorizable diagrams are left in the form of a residual radial integration,
\begin{equation}
  \int_0^\infty dQ^2 \, Q^2 \,
  \Pi^\had(-Q^2) \, f(Q^2,s,t,u)
  \label{eqn:basicint}
\end{equation}
which is calculated numerically once provided with the hadronic vacuum polarization in the space-like region.

One must note, however, that the integral~\eqref{eqn:basicint} requires the knowledge of $\Pi^\had(q^2)$ for any $q^2<0$, while $\Pi^\had$ is experimentally accessible only in a finite range of~$t$. 
Therefore the proposed iterative procedure  will require in any case an extrapolation between the measured region and the high-energy tail close to infinity. Lattice data could also come to the aid in the intermediate region.
Pad\'{e} approximants could be used in this merging procedure. They have been employed, for example, to evaluate the QED vacuum polarization function at four loops and its contribution to the \gmt at five loops~\cite{Baikov:2013ula} knowing the first terms of $\Pi^\lep(q^2)$ in an expansion around $q^2=0,4m^2_\ell, +\infty$. 
It is beyond the scope of this paper to study the impact of such extrapolation and the error that it would introduce in the evaluation of $a_\mu^\mathrm{HLO}$.

The paper is organized as follows. In section~\ref{sec:hyp} we will review the hyperspherical integration method. The master formulae for the evaluation of the irreducible diagrams will be presented in section~\ref{sec:vertex} for the vertex corrections and section~\ref{sec:boxes} for the boxes. In section~\ref{sec:results} we will make a comparison between the traditional dispersive method and the hyperspherical method. Conclusion are drawn in section~\ref{sec:conc}.
The appendix contains an example of a one-loop calculation with the hyperspherical method.

\section{The Hyperspherical Integration Method}
\label{sec:hyp}
In this section we will give a short review of the hyperspherical integration method.
Each of the diagrams in class IV contains an insertion of the SM vacuum polarization tensor with four momentum $q$,
\begin{equation}
  i \Pi^{\mu\nu}(q) = i \Pi(q^2) (g^{\mu\nu}q^2-q^\mu q^\nu) =
  \int d^4x \, e^{iqx} \, \bra{0} T \{j^\mu_{\mathrm{em}}(x) j^\nu_{\mathrm{em}}(0) \} \ket{0} \, ,
\end{equation}
where $j^\mu_{\mathrm{em}}(x) = \sum_f Q_f \bar{\psi}_f(x) \gamma^\mu \psi_f(x)$ is the electromagnetic current and the sum runs over fermions with charges $Q_f$. The weak interactions will be ignored.
Throughout this paper we will always assume $\Pi(q^2)$ to be the renormalized vacuum polarization fulfilling $\Pi(q^2=0)=0$.
In each loop diagram we choose the routing of the loop momentum $q$ in such a way that the momentum flowing through the dressed photon propagator is exactly $q$, so that the loop integral has the following form:
\begin{equation}
  I(p_1,\dots,p_n) = 
\int d^4 q
\frac{\Pi^\had(q^2)}{q^2+ i \varepsilon} \frac{\mathcal{N}(q,p_1,\dots,p_n)}{\mathcal D_1  \cdots \mathcal D_n},
  \label{eqn:oneloop}
\end{equation}
where $n=2 \, (3)$ for the vertex (the box) corrections, $\mathcal D_i = (q+k_i)-m_i^2+i \varepsilon$ are propagator denominators and $k_i$ are linear combination of the external momenta $p_1,\cdots,p_n$.  
The numerator $\mathcal{N}(q,p_1,\dots,p_n)$ is assumed to be a scalar function.
We will work in $D=4$ dimensions. This choice is dictated mainly from the fact that hyperspherical integration of three Feynman propagators, necessary for the boxes, are known only in four-dimensions~\cite{Laporta:1994mb}. More details will be given further on. 

\begin{figure}[htb]
  \centering
  \begin{minipage}[t]{0.45\textwidth}
    \centering
    \includegraphics[width=.8\textwidth]{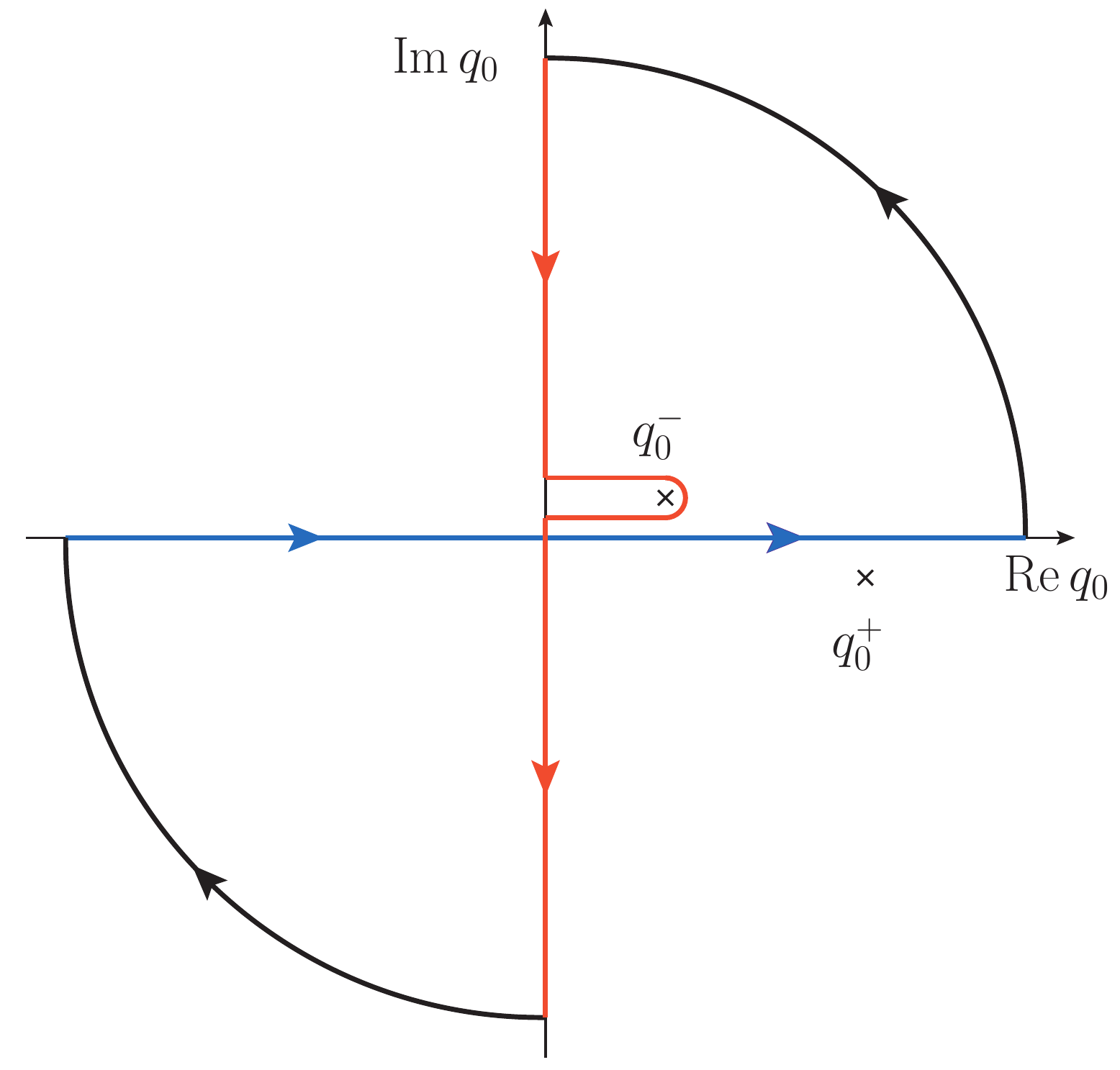}
    \subcaption{}
    \label{fig:wickbefore}
  \end{minipage}
  \begin{minipage}[t]{0.45\textwidth}
    \centering
    \includegraphics[width=.8\textwidth]{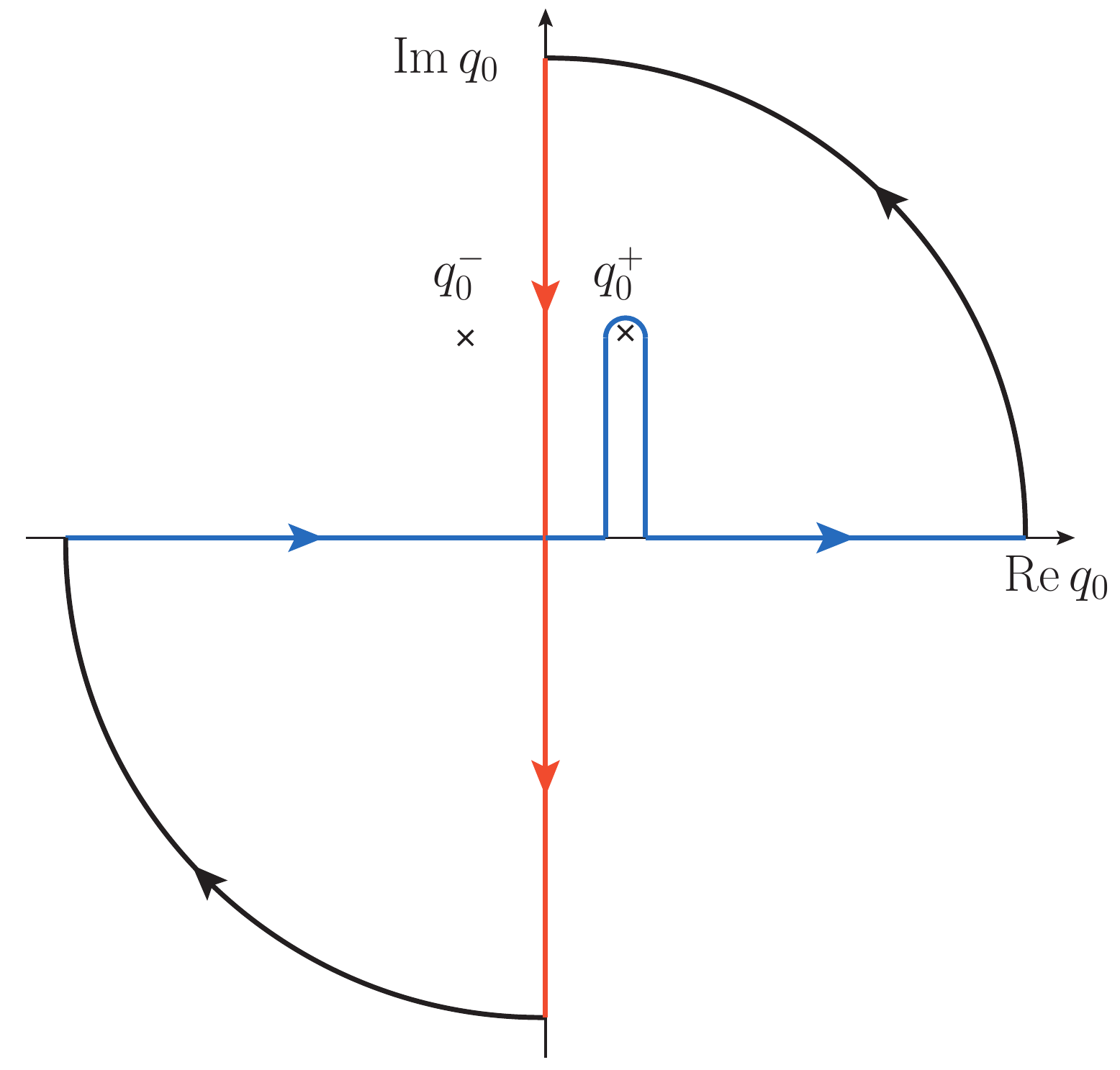}
    \subcaption{}
    \label{fig:wickafter}
  \end{minipage}
  \caption{Integration contour in the complex plane $q_0$ before~(\ref{fig:wickbefore}) and after~(\ref{fig:wickafter}) analytic continuation of the external momenta into the Euclidean region. The Feynman propagator poles $q^\pm_0$ move in the complex plane and the blue and red integration paths must be deformed accordingly.}
\end{figure}

We begin with the analytic continuation of all external momenta into the Euclidean region and with a Wick rotation of the integration contour.
Do we need at this point to add the propagator pole residues after the Wick rotation? 
When we compute a one-loop integral in the traditional way, after introducing Feynman parameters and shifting the loop momentum one is left with a denominator of the form $1/(q^2-\Delta+i \varepsilon)^n$, that has poles at $q_0 = \pm \sqrt{\vec{q} \, ^2+\Delta} \mp i \varepsilon$. Since the poles are in the top-left and bottom-right quadrants of the complex $q_0$ plane, integrating over the real axis is equivalent to integrating over the imaginary axis.
However in the hyperspherical approach, we cannot shift the loop momentum and therefore we are left at the denominator with the product of propagators of the form $1/[(q-p)^2-m^2]$, that has two poles in the $q_0$ complex plane at
\begin{equation}
  q_{0}^\pm = p_0 \pm \sqrt{(\vec q-\vec p \, )^2+m^2} \mp i \varepsilon.
  \label{eqn:poles}
\end{equation}
The two poles are not centered at the origin.
If $p_0$ is sufficiently large, $q_0^-$ lies in the top-right quadrant of the $q_0$ plane. Therefore the integration over the real axis is different from the integration along the imaginary axis: the residue of the pole $q_{0}^-$ must be taken into account also.
Phrased differently, integration over the real axis (in blue in figure~\ref{fig:wickbefore}) is equivalent to the red path in figure~\ref{fig:wickbefore}, which proceeds along the imaginary axis but avoid the $q_0^-$ pole turning around it. 

Analytic continuation of the external momenta into the Euclidean region moves then the location of these poles. For example we can let the energy $p_0$ in~\eqref{eqn:poles} to acquire a phase $e^{i \phi}$ which is then varied from 0 to $\pi/2$. 
In this way, the pole $q_0^-$ moves to the top-left quadrant, while the pole $q_0^+$ to the top-right one. 
No pole should cross the integration contours. Therefore the blue path along the real axis must be deformed because the pole $q_0^+$ moves to the upper side, as shown in figure~\ref{fig:wickafter}, while the path along the imaginary axis becomes straight, in red in figure~\ref{fig:wickafter}, since the pole $q_0^-$ moves to the left. 
So after Wick rotation no pole residue must be included if both internal and external momenta become Euclidean.

Note in addition that $\Pi^\had(q^2)$ has a branch point at the pion threshold $q^2=4m_\pi^2$. In the $q_0$ complex plane it corresponds to two branch points at $q_0 = \pm \sqrt{\vec q \, ^2+ 4m_\pi^2} \mp i \varepsilon$. They are unaffected by the analytic continuation because their position is independent on the external momenta. So they do not interfere with the Wick rotation. 
Furthermore, the vacuum polarization does not introduce any other isolated singularity since its poles in the $q^2$ complex plane, corresponding to unstable resonances, are hidden below the real axis in the unphysical sheet.

Now four-dimensional hyperspherical coordinates can be introduced for the loop momentum $q$: 
\begin{equation*}
   d^4q = i d^4Q = i \frac{Q^2}{2} dQ^2 d\Omega_Q.
\end{equation*}
We will denote with capital and lowercase letters the momenta in the Euclidean and in the Minkowski space, respectively. Note that after Wick rotation the vacuum polarization's argument becomes negative: $\Pi^\had(-Q^2 <0 )$.
The angular dependence of the propagators $\mathcal{D}_i$ in~\eqref{eqn:oneloop} can be made explicit by the expansion
\begin{equation}
  \frac{1}{(Q-P)^2+m^2} 
  = \frac{Z_{Q P}}{| Q || P |} 
  \sum_{n=0}^\infty
  Z_{Q P}^n \, C_n^{(1)} (\hat Q \cdot \hat P),
  \label{eqn:detexpansion}
\end{equation}
where $\hat Q$ and $\hat P$ are unit vectors along the direction of $Q$ and $P$,
\begin{equation}
  Z_{Q P} = \frac{Q^2+P^2+m^2-\lambda^{1/2}(Q^2,P^2,-m^2)}{2| Q|| P|}
  \label{eqn:definitionZ}
\end{equation}
and $\lambda(x,y,z) = x^2+y^2+z^2-2x y -2xz-2yz$ is the K\"{a}ll\'{e}n function.
The Gegenbauer polynomials $C^{(1)}_n(x)$ are an orthogonal basis of functions over the interval $[-1,1]$ with respect to the weight function $\sqrt{1-x^2}$.
This allows us to perform the angular integration using the orthogonality conditions
\begin{gather}
  \int \frac{d\Omega_Q}{2 \pi^2} \,
  C_n^{(1)}(\hat Q \cdot \hat P_i)\, C_m^{(1)}(\hat Q \cdot \hat P_j)
 =
  \frac{\delta_{nm}}{n+1} C_n^{(1)}(\hat P_i \cdot \hat P_j),
  \label{eqn:orthogonalitycondition} \\
   C_n^{(1)}(x)\, C_m^{(1)}(x)
 =
 \sum_{j=0}^{\min(n,m)} C^{(1)}_{m+n-2j} (x).
\end{gather}
Since after  the angular integration the momenta are still space-like, i.e. $p_i^2 <0$, we need eventually to analytically continue back the results to the time-like region.
An example of a one-loop integral calculation with the hyperspherical method is presented in the appendix, where its analytic continuation is also further discussed. 

There is a caveat however: the integral of the product of three Gegenbauer polynomials evaluated at three different $\hat Q \cdot \hat P_i$ is unknown and therefore the angular integration cannot be performed with the method described above. 
This occurs when we calculate the box diagrams: the product of three denominators --- the fourth does not depend on the angles --- would become the product of three Gegenbauer polynomials if the expansion~\eqref{eqn:detexpansion} were employed.
The angular integrals can be evaluated nevertheless by brute force integrating directly with respect to the three hyperspherical angles,
\begin{equation}
  \int d\Omega_Q = 
  \int_0^\pi \sin^2 \theta_1 d\theta_1
  \int_0^\pi \sin \theta_2 d\theta_2
  \int_0^{2\pi} d \phi_3,
\end{equation}
avoiding the expansion~\eqref{eqn:detexpansion}.
The general solution of an integral with three denominators in $D=4$ was given long time ago by Laporta in a not very-well-known article~\cite{Laporta:1994mb}.
This is the main reason why  our calculation is carried out in $D=4$ and not in dimensional regularization, even if the hyperspherical method can be applied to $D$ dimensions as well (see e.g.~\cite{Mastrolia:2016dhn}).

After the hyperspherical integration, the residual $Q^2$ radial integral can be ill-defined because of the bad behaviour of the integrand at infinity --- an ultraviolet (UV) divergence --- or at some finite value of $Q^2$, an infrared (IR) one.  
Vertex corrections are UV divergent but IR finite because $\Pi^\had(0)=0$ regularizes the behaviour of the kernel at the origin. 
Thanks to the on-shell renormalization prescription, the radial integral can be regularized by calculating the vertex together with its counter-term, which is the vertex itself in the limit of zero momentum transfer $ t \to 0$.
The counter term built in this way cancels the original UV divergence pointwise in momentum space~\cite{Levine:1974xh,Roskies:1990ki}.

Vice versa, the boxes contain soft IR divergences but are UV finite. 
A small photon mass $\lambda$ can be introduced to regularize the integral. 
It is even possible to avoid an explicit calculation of IR divergent integrals containing $\Pi^\had(q^2)$ by observing that in the soft limit the box diagrams are proportional to the ``tree-level'' amplitude, i.e.\ the Born amplitude with a dressed photon propagator in figure~\ref{fig:hadLO}. Indeed the soft pole arises when the momentum of the undressed photon goes to zero and the momentum of the dressed one is $t$. This suggests us the possibility to extract the IR poles with the following subtraction: 
\begin{equation}
  \int d^4 q  \, \Pi^\had(q^2) \,   \dots 
  = 
  \int d^4 q  
  \Big[ \Pi^\had(q^2)-\Pi^\had(t) \Big] \,  \dots 
  +\Pi^\had(t)\int d^4 q 
   \dots  \, .
  \label{eqn:irsubtraction}
\end{equation}
The first integral on the r.h.s. of eq.~\eqref{eqn:irsubtraction} is now free of IR divergences and can be evaluated setting $\lambda=0$. 
The soft pole appears only in the second integral of~\eqref{eqn:irsubtraction} that does not contain anymore $\Pi^\had$ and can be calculated analytically with standard methods.

The last technical ingredient to discuss is how to perform the angular integration when the loop momentum $q$ appears also at the numerator.
One occurrence of the scalar product $k_i \cdot q$ can be always removed against one of the propagators $\mathcal D_i$ by writing $2 k_i \cdot q = \mathcal D_i-q^2+m^2-k_i^2$. 
Additional $k_i \cdot q$ in the numerator can be further simplified  using the technique described in the appendix of ref.~\cite{Roskies:1990ki}.
The one-denominator case is straightforward:
\begin{equation}
  \int \frac{d\Omega_Q}{2\pi^2} \frac{k_i \cdot q}{\mathcal D_j} = 
 - \int \frac{d\Omega_Q}{2\pi^2} 
   \frac{|K_i||Q| }{\mathcal D_j} \hat K_i \cdot \hat Q
  = -\int \frac{d\Omega_Q}{2\pi^2} 
   \frac{|K_i||Q|}{\mathcal D_j}
   \frac{C^{(1)}_1( \hat K_i \cdot \hat Q)}{2},
 \end{equation}
given that $C_1^{(1)}(x) = 2 x$. The angular integral is performed by expanding the denominator and using the orthogonality condition~\eqref{eqn:orthogonalitycondition}.
For the two-propagator case we write
\begin{equation}
  \int \frac{d\Omega_Q}{2\pi^2}
  \frac{K_i \cdot Q}{\D_j \D_k} =
  K_i^\mu I^\mu, 
  \quad \text{with }
  I^\mu = 
  \int \frac{d\Omega_Q}{2\pi^2}
  \frac{Q^\mu}{\D_j \D_k}.
  \label{eqn:Imu}
\end{equation}
The term $I_\mu$ must be a linear combination of the Euclidean vectors $K_j$ and $K_k$ that appear at the denominator with scalar coefficients.
Introducing two orthonormal vectors $\hat e^\mu_1$ and $\hat e^\mu_2$ in the two-dimensional space spanned by $K_j$ and $K_k$,
\begin{align}
  \hat e^\mu_1 & = \frac{K_j^\mu}{|K_j|}, &
  \hat e^\mu_2 &= 
  \frac{K_k^\mu- K_j^\mu (K_j\cdot K_k) / K_j^2}{\sqrt{K_k^2-(K_j \cdot K_k)^2 / K_j^2}},
\end{align}
we can replace the loop momentum $Q^\mu$ in~\eqref{eqn:Imu} with its projection onto the space spanned by $K_j$ and $K_k$:
\begin{equation}
  Q^\mu \to  \hat e_1^\mu (\hat e_1 \cdot Q) + \hat e_2^\mu (\hat e_2 \cdot Q).
\end{equation}
Now the integral $I^\mu$ contains, through the scalar product $\hat e_i \cdot Q$, terms like $K_j \cdot Q /\D_j$ or $K_k \cdot Q /\D_k$, which can be simplified as before, leading to an integrand without $Q$ in the numerator.

\begin{figure}[t]
\centering
\includegraphics[width=.3\textwidth]{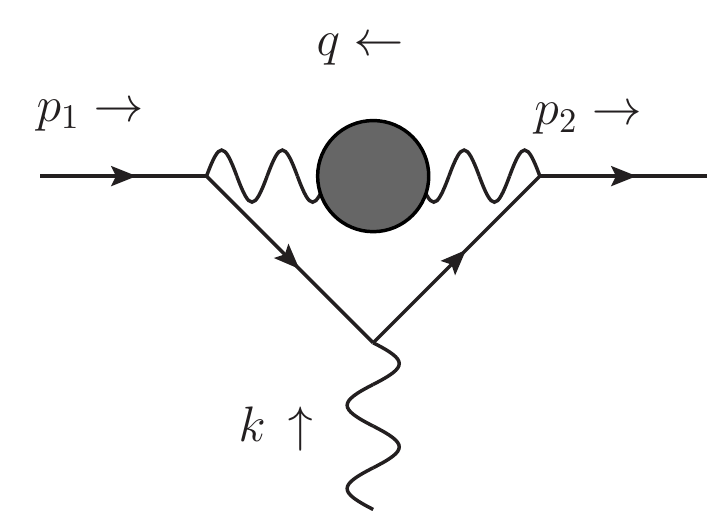}
\caption{The leading contribution of the hadronic vacuum polarization to the QED vertex.}
\label{fig:HLO}
\end{figure}
\section{The Vertex Corrections \label{sec:vertex}}
Having introduced the hyperspherical method, we can now apply it to calculate the hadronic vacuum polarization contribution to the QED form factors which can be used in a second stage to calculate the irreducible vertex corrections in class IV.  
The 1PI amplitude $\Gamma_\mu(k)$ describing the interaction between a photon and the initial and final states of an on-shell lepton $\ell$, with four-momenta $p_1$ and $p_2$, respectively, can be written in terms of the Dirac and Pauli form factors $F_1$ and $F_2$: 
\begin{equation}
\Gamma_\mu^\ell(k)=
 \gamma_\mu F_1^\ell(k^2) +i \frac{\sigma_{\mu\nu}k^\nu}{2m_\ell} F_2^\ell(k^2),
\end{equation}
where $\sigma^{\mu\nu}=\frac{i}{2}[\gamma^\mu,\gamma^\nu]$, $m_\ell$ is the lepton mass  and $k=p_2-p_1$ is the incoming four-momentum of the off-shell photon.
Let us call $F_{1}^{\ell \, \had}(k^2)$ and $F_2^{\ell \,\had}(k^2)$ the leading contribution of the hadronic vacuum polarization to the form factors $F_1^\ell$ and $F_2^\ell$ given by the two-loop diagram in figure~\ref{fig:HLO}.
The vertex corrections in class IV can be expressed in term of $F_1^{\ell \, \had} (k^2)$ and $F_2^{\ell \, \had}(k^2)$, with $\ell=e,\mu$.
Since there the role of the off-shell photon with momentum $k$ is played by the photon in $t$-channel, we will identify $k^2$ with the Mandelstam variable $t<0$, and we will restrict the calculation to the region $t<0$.

The form factors $F_1^{\ell \, \had} (k^2)$ and $F_2^{\ell \, \had}(k^2)$ are extracted from the amplitude with the projector technique~\cite{Barbieri:1972as} and the loop integral calculated with the hyperspherical method. 
The final expression for the form factors can be cast in the form
\begin{equation}
  F_{i}^{\ell \, \had}( t ) = 
-\frac{\alpha}{\pi} 
\int_0^1 dx  \,
\Pi^\had \left( \frac{m^2_\ell x^2}{x-1} \right) 
 \, f_{i}\left(x,\frac{t}{m_\ell^2}\right),
\label{eqn:ffintrep}
\end{equation}
where $i=1,2$ and $x$ is related to the radial variable $Q^2$ by
\begin{equation}
q^2 = -Q^2 = - \frac{m^2_\ell x^2}{1-x}.
\end{equation}
The angular integration gives a well-behaved integrand $f_2$, while we need to renormalize $F_1^\mathrm{had}$. 
We impose the on-shell renormalization condition $F_1^\had(0) = 0$ by subtracting, as a counter term, the integrand itself in the limit $k^2 \to 0$.
The final expressions for the kernel functions appearing in the renormalized form factors are:
\begin{align}
f_1(x,y) &= 
\frac{3 x^3-4 x^2+4}{4(1-x) x}
+\frac{2-x}{1-x} \Bigg\{
\frac{6 x^2}{(4-y)^2 (x-1)} 
+\frac{x^2-6 x+4}{2 (4-y) (x-1)} \notag \\
&+\left[ 
\frac{\left(x^2+8 x-8\right) x}{(4-y) (1-x)^2}
-\frac{12 x^3}{(4-y)^2(1-x)^2}
+\frac{4-y}{x}
+\frac{2 \left(x^2+x-1\right)}{(1-x) x}
\right] \notag \\
&\times \left.\frac{1}{\sqrt{y(y-4)}}
\arctanh \left( 
\frac{(1-x)\sqrt{y \, (y-4)}}{2x+y-4-yx}
 \right)
\right\rbrace, \label{eqn:f1} \\
f_2(x,y) &=\frac{2-x}{1-x} \Bigg\{
\frac{6x^2}{(4-y)^2(1-x)}
+\frac{2-x}{4-y}  
+\left[
\frac{2x}{(4-y)(1-x)}
+\frac{3x^3}{(4-y)^2(1-x)^2}
\right]
\notag \\
&\times \left. \frac{4}{\sqrt{y(y-4)}}
\arctanh \left( 
\frac{(1-x)\sqrt{y \,(y-4)}}{2x+y-4-yx}
 \right)
\right\rbrace \, ,
\label{eqn:f2}
\end{align}
valid in the scattering region $t<0$.
The inverse hyperbolic tangent appearing in~\eqref{eqn:f1} and~\eqref{eqn:f2} are always real-valued if $0<x<1$ and $y<0$. 

By taking the limit $t\to 0$ in $F_2^{\ell \, \had}$, we recover the space-like formula for $a_\mu^\mathrm{HLO}$ in eq.~\eqref{eqn:aHLOspace}, which is usually derived by applying twice the dispersion relation. Our calculation shows that eq.~\eqref{eqn:aHLOspace} can be obtained directly, without making use of the dispersion relation (it was proven already in~\cite{Blum:2002ii}).
Moreover by substituting $\Pi(q^2) \to \!\! -1$ in~\eqref{eqn:ffintrep}~and performing the integration analytically, we reproduce the Pauli form factor at one-loop~\cite{Barbieri:1972as}:
{
\setlength{\abovedisplayskip}{3pt}
\setlength{\belowdisplayskip}{3pt}
\begin{equation}
F_2(k^2) = \frac{\alpha}{\pi}
\frac{\xi \log \xi}{\xi^2-1},
\end{equation}
}
where $\xi$ is the Landau variable $t/m_\ell^2 = - (1-\xi)^2/\xi$. %
The same check cannot be done straightforwardly for the Dirac form factor because $f_1(x,y)$ is not integrable anymore in $x=0$ if we set $\Pi^\had(q^2)=-1$, while we assumed $\Pi^\had(0)=0$. To reproduce $F_1(k^2)$ at one loop, we can substitute in~\eqref{eqn:ffintrep}
\begin{equation}
\Pi(q^2) \to -\frac{q^2}{q^2-\lambda^2},
\end{equation}  
which corresponds in figure~\ref{fig:HLO} to the exchange of the dressed photon with an undressed one with fictitious mass $\lambda$. 
The integral~\eqref{eqn:ffintrep} is now finite and the integration can be done analytically. Keeping terms that do not vanish in the limit $\lambda \to 0$ we correctly recover the known result~\cite{Barbieri:1972as}:
\begin{align}
F_1(k^2)  &=  \left(\frac{\alpha}{\pi}\right) 
\Bigg\lbrace
\log \left( \frac{\lambda}{m}\right) 
\left(\frac{\xi ^2+1 }{\xi^2-1}\log (\xi )-1\right) 
+\frac{3 \xi ^2+2 \xi +3 }{4 \left(\xi ^2-1\right)}\log (\xi )-1  \notag \\
&+\frac{1+\xi^2}{1-\xi ^2}
\left[\text{Li}_2(-\xi )  
-\frac{\log ^2(\xi )}{4} 
+\frac{\pi ^2}{12}+\log (\xi ) \log (\xi +1)\right] \Bigg\rbrace.
\end{align}

\begin{figure}[th]
  \centering
  \begin{minipage}[t]{0.45\textwidth}
    \centering
    \includegraphics[width=0.8\textwidth]{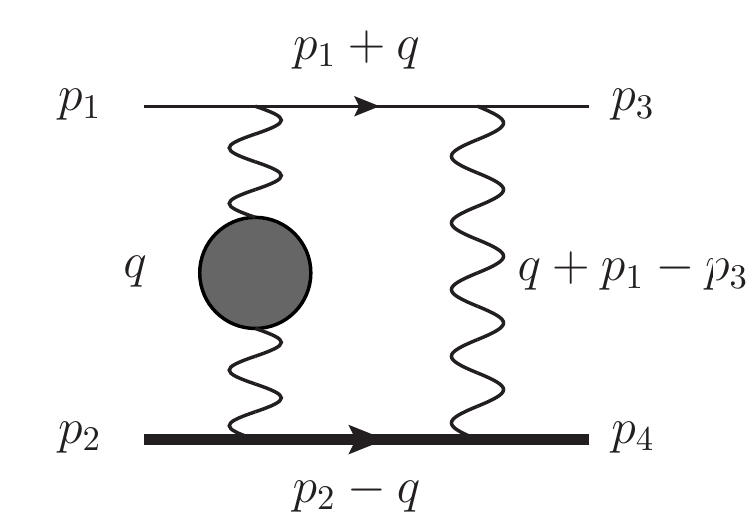}
    \subcaption{}\label{fig:box}
  \end{minipage}
  \begin{minipage}[t]{0.45\textwidth}
    \centering
    \includegraphics[width=0.8\textwidth]{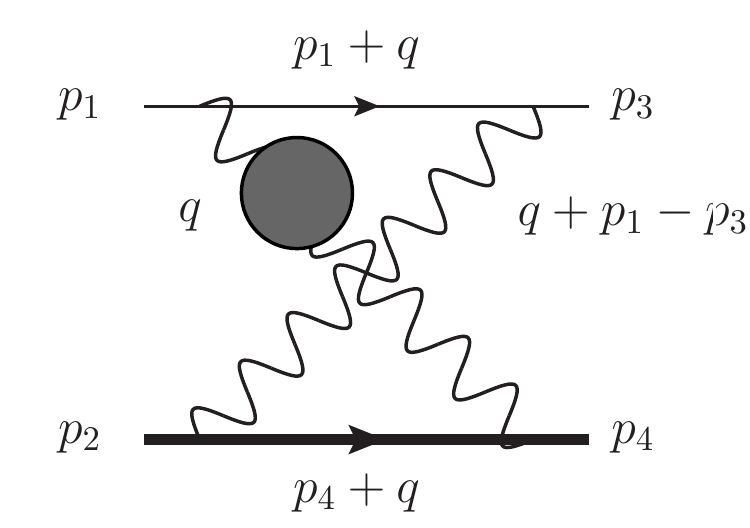}
    \subcaption{}\label{fig:boxcross}
  \end{minipage}
  \caption{Irreducible hadronic box diagrams contributing to $\mu$-$e$ scattering at NNLO. Muon and electron lines are depicted with thick and  thin lines.
  Two additional diagrams, with the vacuum polarization in the other photon propagator,  must be considered also.}
  \label{fig:boxes}
\end{figure}
\section{The Boxes \label{sec:boxes}}
We can now turn our attention to the box diagrams.
There are two topologies to take into account: the uncrossed two photon exchange in figure~\ref{fig:box} and the crossed one in figure~\ref{fig:boxcross}.
They are related by the crossing $s \leftrightarrow u$ plus an overall minus sign.
Both photons can be dressed with the hadronic vacuum polarization. 
The diagram with the same topology but with dressed and undressed photon exchanged can be obtained by replacing the initial state momenta with the final state ones and vice versa.
Therefore the contribution of the two diagrams to the unpolarized cross section is the same since no crossing of the Mandelstam variables occurs.  
Let us fix the notation for the process $e^- \mu^- \to e^- \mu^-$. We choose the following set of propagators:
\begin{align}
  \mathcal D_0 & = q^2, & 
  \mathcal D_1 & = (q+p_1)^2-m^2, \notag \\ 
  \mathcal D_2 & = (q+p_1-p_3)^2, & 
  \mathcal D_3 & = (q-p_2)^2-M^2, \notag \\
  \mathcal D_4 & = (q+p_4)^2-M^2, 
  \label{eqn:Dens}
\end{align}
where $m^2$ ($M^2$), $p_1$ and $p_3$ ($p_2$ and $p_4$) are the mass, the initial state and the final state momentum of the electron (the muon). The Mandelstam variables
\begin{align}
  s &= (p_1+p_2)^2 = (p_3+p_4)^2,\\
  t &= (p_1-p_3)^2 = (p_2-p_4)^2,\\
  u &= (p_1-p_4)^2 = (p_2-p_3)^2 ,  
  \label{eqn:Mandelstam}
\end{align}
satisfy $s+t+u=2m^2+2M^2$, with the physical requirements 
\begin{gather}
    (m+M)^2 < s, \\
   -\frac{\lambda(s,M^2,m^2)}{s} < t < 0, \\
   2m^2+2M^2-s < u < \frac{(M^2-m^2)^2}{s}.
 \end{gather}

We work at the level of interferences between the boxes and the Born amplitude which provide us with the scalar numerators $\mathcal{N}$ in eq.~\eqref{eqn:oneloop}.
After algebraic manipulation, these interferences are written as linear combinations of loop integrals that can be evaluated one by one via the hyperspherical method. 
Each of the box diagrams requires the evaluation of 14 master integrals that have the following form:
\begin{equation}
  I = \frac{1}{i \pi^2} \int d^4 q \, \Pi^\had (q^2) \, \dots 
  = \int_0^{+\infty} dQ^2 \, Q^2 \, \Pi^\had (-Q^2) \, 
  \Big\langle \dots \Big\rangle \, . 
  \label{eqn:defI}
\end{equation}
The kernel functions denoted by 
\begin{equation}
  \Big\langle \dots \Big\rangle =
  \int \frac{d\Omega_Q}{2\pi^2}
  \dots \, \Bigg\vert_{p_i^2 \to m_i^2} \,,
  \label{eqn:angledef}
\end{equation}
arise from the angular integration of their arguments followed by analytic continuation of the external momenta back to the physical region.
The solutions of the angular integration (for Euclidean momenta) are taken from the results in refs.~\cite{Knecht:2001qf,Jegerlehner:2009ry,Laporta:1994mb}.
The necessary angular integrals are the following: 
\begin{align}
  \left\langle
  \frac{1}{\D_1}
  \right\rangle &=
  \frac{1}{2 m^2}
  \left(1- \sqrt{1+\frac{4m^2}{Q^2}} \right),
  \label{eqn:d1} \\[7pt]
  \left\langle
  \frac{1}{\D_2}
  \right\rangle &=
  \frac{\theta(-Q^2-t)}{t} - \frac{\theta(Q^2+t)}{Q^2},
  \label{eqn:d2} \\[7pt]
  \left\langle
  \frac{1}{\D_3}
  \right\rangle =
  \left\langle
  \frac{1}{\D_4}
  \right\rangle &=
  \frac{1}{2 M^2}
  \left(1- \sqrt{1+\frac{4M^2}{Q^2}} \right),
  \label{eqn:d3} 
   \\[7pt]
  \left\langle
  \frac{1}{\D_1 \D_2}
  \right\rangle &=
  -\frac{1}{Q^2 \sqrt{t(t-4m^2)}} 
  \left[
    L(z_1)+ 2 \theta(Q^2+t) \, L\left( \frac{1}{z_2}\right)
  \right],
  \label{eqn:d1d2} \\[7pt]
  \left\langle
  \frac{1}{\D_2 \D_3}
  \right\rangle  =
  \left\langle
  \frac{1}{\D_2 \D_4}
  \right\rangle &=
  -\frac{1}{Q^2 \sqrt{t(t-4M^2)}} 
  \left[
   L(z_3)+2\theta(Q^2+t) \, L\left( \frac{1}{z_4}\right)  \right],
  \label{eqn:d2d3} \\[7pt]
  \left\langle
  \frac{1}{\D_1 \D_3}
  \right\rangle &=
  +\frac{1}{Q^2 \lambda^{1/2}(s,m^2,M^2)} 
  \Big[
    L(z_5)+L(z_6)-L(z_7)
  \Big],
  \label{eqn:d1d3}\\[7pt]
  \left\langle
  \frac{1}{\D_1 \D_4}
  \right\rangle &=
  +\frac{1}{Q^2 \lambda^{1/2}(u,m^2,M^2)} 
  \notag \\
  &\times \mathrm{Re} \, \Big[
  L(z_8)- L(z_9)+L(\sgn(u+M^2-m^2) z_{10})
  \Big] \, ,
  \label{eqn:d1d4}
  \\[7pt]
  \left\langle
  \frac{1}{ \D_1 \D_2 \D_3}
  \right\rangle &= -
  \frac{1}{Q^2 \, |Q^2+t| \, \lambda^{1/2}(s,m^2,M^2)} \notag \\[7pt]
  &\times\Big[
    L\big(\sgn(Q^2+t) \,  z_5\big)
    +L\big(\sgn(Q^2+t) \, z_6\big)-L(z_7)
  \Big],
  \label{eqn:d1d2d3}
  \\[7pt]
  \left\langle
  \frac{1}{ \D_1 \D_2 \D_4}
  \right\rangle &= -
  \frac{1}{Q^2 \, |Q^2+t| \, \lambda^{1/2}(u,m^2,M^2)} 
  \mathrm{Re} \, \Bigg[
    -L\Big(\sgn(Q^2+t) \,  z_9\Big)
    \notag \\[7pt]
  &+L\Big(\sgn(Q^2+t) \, \sgn(u+M^2-m^2) \, z_{10}\Big)
  +L\Big(z_8\Big )
  \Bigg],
  \label{eqn:d1d2d4}
\end{align}
where
\begin{align}
  z_1 &=\frac{\sqrt{\uprule Q^2 (Q^2+4m^2)}
  \sqrt{\uprule t(t-4m^2)}}{2m^2 t -2m^2 Q^2 +Q^2 t} \,, &
  z_2 &= \sqrt{1-\frac{4m^2}{t}} \, , \\[7pt]
  z_3 &= \frac{\sqrt{\uprule Q^2 (Q^2+4M^2)} 
  \sqrt{\uprule t(t-4M^2)}}{2M^2 t -2M^2 Q^2 +Q^2 t} \, ,&
  z_4 &= \sqrt{1-\frac{4M^2}{t}} \, , \\[7pt]
  z_5 &= 
  \sqrt{1+\frac{4m^2}{Q^2}} \sqrt{1-\frac{4s m^2}{(s-M^2+m^2)^2}} \, , &
  z_6 &= 
  \sqrt{1+\frac{4M^2}{Q^2}} \sqrt{1-\frac{4s M^2}{(s-m^2+M^2)^2}} \, , \\[7pt]
  z_7 &=\sqrt{1-\frac{4m^2M^2}{(s-M^2-m^2)^2}} \, , &
  z_8 &= \sqrt{1-\frac{4m^2M^2}{(u-M^2-m^2)^2}} \, , \\[7pt]
  z_9 &=\sqrt{1+\frac{4m^2}{Q^2}} \sqrt{1-\frac{4u m^2}{(u-M^2+m^2)^2}}\, , &
  z_{10} &=  \sqrt{1+\frac{4M^2}{Q^2}} \sqrt{1-\frac{4u M^2}{(u-m^2+M^2)^2}} \, ,
\end{align}
and
\begin{equation}
  L(z) \equiv \frac{1}{2} \log 
  \left( \frac{1+z}{1-z}  \right) \, .
  \label{eqn:defL}
\end{equation}
The function $L(z)$ has branch cuts on the real axis in $]-\infty,-1]$ and $[1,+\infty[$, with $\mathrm{Im} \, L(x) = i \pi/2$ if $x$ is real and $|x|>1$. 
The formula~\eqref{eqn:defL} is often used to define the inverse hyperbolic tangent via $\arctanh(x) \equiv L(x)$. However such identity must be taken with care since some program languages, like for example Mathematica or the GNU Scientific Library, define $\arctanh(z) \equiv \frac{1}{2}\log(1+z)-\frac{1}{2}\log(1-z)$, that assigns a negative imaginary part to the function if $x>1$: $\mathrm{Im} \, \arctanh(x) = - i \pi/2$.
 
At this point we would like to comment on the analytic continuation that we performed in eqs.~(\ref{eqn:d1}-\ref{eqn:d1d2d4}).
The general expressions for the angular integrals in~\cite{Knecht:2001qf,Jegerlehner:2009ry,Laporta:1994mb} are written in terms of squared Euclidean momenta fulfilling $P_i^2 > 0$ and $| \hat P_i \cdot  \hat P_j | <1$.
They must be continued to the on-shell conditions $P_i^2=-m_i^2$ and $(P_1-P_2)^2 = -(s+i \varepsilon), (P_1-P_3)^2 = - t, (P_1-P_4)^2 = -u $. 
Analytic continuation affects the whole radial integral~\eqref{eqn:defI}, not only the angular integration result. Indeed it is necessary to check if any singularity crosses, in the $Q^2$ complex plane, the integration path along the positive real axis when the $P_i^2$ are continued from positive to negative values. 
This check is carried out explicitly for the one-loop example in the appendix.

Note that when making use of the results in ref.~\cite{Knecht:2001qf,Jegerlehner:2009ry,Laporta:1994mb} one just need to identify the correct side of the branch cut of the functions $L(z)$ after setting $P_i^2=-m_i^2$. 
This can be achieved by imposing the correct analyticity structure dictated by unitarity.
Indeed after substitution of $P_i^2= - m_i^2$ and $(P_1-P_2)^2 = -s$ etc., all the square roots in front of~(\ref{eqn:d1d2}-\ref{eqn:d1d2d4}) become real and positive and all arguments $z_i$ are real too. 
The vacuum polarization function is real in the space-like region, therefore the imaginary part of $L(z_i)$ is the only one that must be fixed.
Let us consider for example the integrals~\eqref{eqn:d1d2} and~\eqref{eqn:d2d3} which are obtained by pinching the denominators $\D_3$ or $\D_1$ in the box.
These integrals depend on $t<0$ and therefore, since they are evaluated below the lepton-pair threshold, their imaginary part must be equal to zero. 
The imaginary parts of the $L(z_i)$ can be chosen accordingly with this constraint.
Similar arguments apply to the integrals for the crossed box diagrams that depend on the Mandelstam variable $u$.

In addition to the formulae in~(\ref{eqn:d1}-\ref{eqn:d1d2d4}), angular integrals with scalar product $q \cdot p_i $ at the numerator are necessary.  
Thanks to the reduction technique outlined at the end of section~\ref{sec:hyp}, they can be written as a linear combination of the integrals~(\ref{eqn:d1}-\ref{eqn:d1d4}):
\begin{align}
  \left\langle
  \frac{q\cdot p_1 }{\D_0 \D_2 \D_3} 
  \right\rangle &=
  - \frac{1}{2(4M^2-t)}
  \Bigg\lbrace 
  \frac{s+M^2-m^2}{Q^2}
  \left\langle
  \frac{1}{ \D_3} 
  \right\rangle 
  +\frac{2 m^2 + 2 M^2 - 2 s - t}{Q^2}
  \left\langle
  \frac{1}{ \D_2} 
  \right\rangle  \notag \\[7pt]
  &-\Bigg[
   s+t-m^2 - 3 M^2 
   +\frac{t (s+M^2-m^2)}{Q^2} 
  \Bigg]
  \left\langle
  \frac{1}{ \D_2 \D_3} 
  \right\rangle
  \Bigg\rbrace,
  \label{eqn:qp1overd2d3} \\[7pt]
  \left\langle
  \frac{q\cdot p_1 }{ \D_0 \D_2 \D_4} 
  \right\rangle &=
  -\frac{1}{2(4M^2-t)}
  \Bigg\lbrace 
  \frac{u+M^2-m^2}{Q^2}
  \left\langle
  \frac{1}{ \D_4} 
  \right\rangle 
  +\frac{2 m^2 + 2 M^2 - 2 u - t}{Q^2}
  \left\langle
  \frac{1}{ \D_2} 
  \right\rangle  \notag \\[7pt]
  &-\Bigg[
   u+t-m^2 - 3 M^2 
   +\frac{t (u+M^2-m^2)}{Q^2}
  \Bigg]
  \left\langle
  \frac{1}{ \D_2 \D_4} 
  \right\rangle
  \Bigg\rbrace,
  \label{eqn:qp1overd2d4}
  \\[7pt]
  \left\langle
  \frac{q\cdot p_2 }{\D_0 \D_1 \D_2} 
  \right\rangle &=
  +\frac{1}{2(4m^2-t)}
  \Bigg\lbrace 
  \frac{s+m^2-M^2}{Q^2}
  \left\langle
  \frac{1}{ \D_1} 
  \right\rangle 
  +\frac{2 m^2 + 2 M^2 - 2 s - t}{Q^2}
  \left\langle
  \frac{1}{ \D_2} 
  \right\rangle \notag \\[7pt]
  &-\Bigg[
   s+t-M^2 - 3 m^2
   +\frac{t (s+m^2-M^2) }{Q^2}
  \Bigg]
  \left\langle
  \frac{1}{ \D_1 \D_2} 
  \right\rangle
  \Bigg\rbrace,
  \label{eqn:qp2overd1d2} \\[7pt]
  \left\langle
  \frac{q\cdot p_4 }{ \D_0 \D_1 \D_2} 
  \right\rangle &=
  -\frac{1}{2(4m^2-t)}
  \Bigg\lbrace 
  \frac{u+m^2-M^2}{Q^2}
  \left\langle
  \frac{1}{ \D_1} 
  \right\rangle 
  +\frac{2 m^2 + 2 M^2 - 2 u - t}{Q^2}
  \left\langle
  \frac{1}{ \D_2} 
  \right\rangle \notag \\[7pt]
  &-\Bigg[
   u+t-M^2 - 3 m^2
   +\frac{t (u+m^2-M^2) }{Q^2}
  \Bigg]
  \left\langle
  \frac{1}{ \D_1 \D_2} 
  \right\rangle
  \Bigg\rbrace,
  \label{eqn:qp4overd1d2} 
  \\[7pt]
  \left\langle
  \frac{q\cdot (p_1-p_3) }{ \D_0 \D_1 \D_3} 
  \right\rangle &=
  \frac{t}{2 \lambda(s,M^2,m^2)} \notag \\
  &\times \Bigg[ 
  2 s 
  \left\langle
  \frac{1}{ \D_1 \D_3} 
  \right\rangle
  +\frac{s+m^2-M^2}{Q^2} 
  \left\langle
  \frac{1}{ \D_1 } 
  \right\rangle 
  +\frac{s+M^2-m^2}{Q^2}
  \left\langle
  \frac{1}{ \D_3 } 
  \right\rangle
  \Bigg],
  \label{eqn:qp13overd1d3} \\
  \left\langle
  \frac{q\cdot (p_1-p_3) }{\D_0 \D_1 \D_4} 
  \right\rangle &=
  \frac{t}{2 \lambda(u,M^2,m^2)} \notag \\
  &\times \Bigg[ 
  2 u 
  \left\langle
  \frac{1}{ \D_1 \D_4} 
  \right\rangle
  +\frac{u+m^2-M^2}{Q^2}
  \left\langle
  \frac{1}{ \D_1 } 
  \right\rangle 
  +\frac{u+M^2-m^2}{Q^2}
  \left\langle
  \frac{1}{ \D_4 } 
  \right\rangle
  \Bigg].
  \label{eqn:qp13overd1d4} 
\end{align}

Up to this point, we have given an account of the angular integration solutions.
We can now introduce the explicit expressions of the radial master integrals that must be evaluated numerically once provided with the hadronic vacuum polarization at negative $q^2$.
The integrals  are the following:
\begin{align}
  I_{0ij} &=
  \int dQ^2 \,Q^2 \, \Pi^\had(-Q^2)
  \left\langle
  \frac{1}{\D_0 \D_i \D_j } 
  \right\rangle \, , \\
  I_{jk}^i &=
  \int dQ^2 \,Q^2 \, \Pi^\had(-Q^2)
  \left\langle
  \frac{q \cdot p_i }{\D_0 \D_j \D_k } 
  \right\rangle \, , \\
   I_{ijk} &=
   \int dQ^2 \,Q^2 \,  \Pi^\had(-Q^2) 
  \left\langle
  \frac{1}{\D_i \D_j \D_k }  
  \right\rangle  \, ,
  \label{eqn:Iijk}\\
   I_{0ijk} &=
   \int dQ^2 \,Q^2 \,  \Pi^\had(-Q^2) 
  \left\langle
  \frac{1}{\D_0 \D_i \D_j \D_k }  
  \right\rangle  \, ,
  \label{eqn:I0ijk}
\end{align}
with $i,j,k=1,2,3$ ($i,j,k = 1,2,4$) for the uncrossed box in figure~\ref{fig:box} (the crossed box in figure~\ref{fig:boxcross}) and $i\neq j \neq k$.
The denominator $\D_0=q^2=-Q^2$ does not depend on the angles, so we can assemble the kernel functions from the results in~(\ref{eqn:d1}-\ref{eqn:d1d2d4}) straightforwardly. 
The following integrals must be considered as well:
\begin{align}
  I_{\Delta 0ik} &=
  \int dQ^2 \,Q^2 \,  \Pi^\had(-Q^2) 
  \left\langle
  \frac{1}{\D_0 \D_i}
  -\frac{\D_0}{\D_1 \D_2 \D_k }  
  \right\rangle  
  \label{eqn:IDelta0ik},\\
  I_{\Delta ijk} &=
  \int dQ^2 \,Q^2 \, \Pi^\had(-Q^2) 
  \left\langle
  \frac{1}{\D_i \D_j}
  -\frac{\D_0}{\D_1 \D_2 \D_k }  
  \right\rangle  .
  \label{eqn:IDeltaijk}
\end{align}
with $i\neq j$, $i,j = 1,2,3 \, (1,2,4)$ and $k=3 \, (4)$ for the uncrossed box (the crossed box).  In eq.~\eqref{eqn:IDelta0ik} and~\eqref{eqn:IDeltaijk} the kernel functions contain two terms, each of them gives a UV divergent integral if taken alone.
To avoid the introduction of an explicit UV regulator, which eventually cancels out in the final result, we take their difference to obtain a UV finite integral.

In eqs.~(\ref{eqn:Iijk}-\ref{eqn:IDeltaijk}), the $1/|Q^2+t|$ pole in the functions~\eqref{eqn:d1d2d3} and~\eqref{eqn:d1d2d4} yields a singular integral that corresponds to the soft IR divergence arising when the undressed photon becomes soft.
Note on the contrary that the $1/Q^2$ pole does not lead to a singular integral since the kernel behaviour is smoothed at $Q^2 \to 0 $ by the renormalized vacuum polarization.
The IR singularity can be regularized by introducing a photon mass $\lambda$ for the undressed photon. However if we perform the subtraction~\eqref{eqn:irsubtraction} for each integral,
\begin{align}
   I_{ijk} &=
   \int dQ^2 \,Q^2 \, \Big[ \Pi^\had(-Q^2) - \Pi^\had(t) \Big]
  \left\langle
  \frac{1}{\D_i \D_j \D_k }  
  \right\rangle \notag \\ 
   &+ \Pi^\had(t)
  \int \frac{d^4 q}{i\pi^2} \frac{1}{\D_i \D_j \D_k} \, ,
  \label{eqn:Iijksub}\\[7pt]
   I_{0ijk} &=
   \int dQ^2 \,Q^2 \, \Bigg[ \Pi^\had(-Q^2) - \frac{2 Q^2}{Q^2+|t|}
   \Pi^\had(t) \Bigg]
  \left\langle
  \frac{1}{\D_0 \D_i \D_j \D_k }  
  \right\rangle  \notag \\
  & + 2 \, \Pi^\had(t)
  \int \frac{d^4 q}{i\pi^2} \frac{1}{(q^2-|t|) \D_i \D_j \D_k} \, ,
  \label{eqn:I0ijksub} 
\end{align}
\begin{align}
  I_{\Delta 0ik} &=
  \int dQ^2 \,Q^2 \, \Big[ \Pi^\had(-Q^2) - \Pi^\had(t) \Big]
  \left\langle
  \frac{1}{\D_0 \D_i}
  -\frac{\D_0}{\D_1 \D_2 \D_k }  
  \right\rangle  \notag \\
   &+ \Pi^\had(t)
  \int \frac{d^4 q}{i\pi^2} 
  \left(\frac{1}{\D_0 \D_i}
  -\frac{\D_0}{\D_1 \D_2 \D_k} \right) ,
  \label{eqn:IDelta0iksub}\\[7pt]
  I_{\Delta ijk} &=
  \int dQ^2 \,Q^2 \, \Big[ \Pi^\had(-Q^2) - \Pi^\had(t) \Big]
  \left\langle
  \frac{1}{\D_i \D_j}
  -\frac{\D_0}{\D_1 \D_2 \D_k }  
  \right\rangle  \notag \\
   &+ \Pi^\had(t)
  \int \frac{d^4 q}{i\pi^2} 
  \left(\frac{1}{\D_i \D_j}
  -\frac{\D_0}{\D_1 \D_2 \D_k} \right) ,
   \label{eqn:IDeltaijksub}
\end{align}
we can still employ the formulae presented before to build the kernel functions.
Indeed, in eqs.~(\ref{eqn:Iijksub}-\ref{eqn:IDeltaijksub}) the first integral is now free of IR divergences because the factor $\Pi^\had(-Q^2)-\Pi^\had(t)$ compensates the $|Q^2+t|$ at the denominator. 
Therefore we can set $\lambda=0$ in this first term and use our results for the angular integrals.
The soft pole appears only in the second terms where standard techniques can be employed for the evaluation of the integrals since $\Pi^\had$ does not depend anymore on the loop momentum $q$.

Note that the simple subtraction~\eqref{eqn:irsubtraction} does not work for $I_{0ijk}$ in~\eqref{eqn:I0ijk}. The kernel has a  $1/Q^2$ pole compensated at $Q^2=0$ by $\Pi^\had(-Q^2)$ but not by a constant term like $\Pi^\had(t)$, whereas the factor $2 \, Q^2/(Q^2+|t|)$ vanishes in the $Q^2 \to 0$ limit while it gives one when $Q^2 \to |t|$.

\section{Dispersive vs Hyperspherical Method \label{sec:results}}

With the formulae for the QED form factors and the boxes in our hand, we can now make a numerical comparison between the standard dispersive approach and the hyperspherical method.
The comparison can be done not only with the hadronic vacuum polarization $\Pi^\had(q^2)$, but also with the well-known analytic expression for $\Pi^\lep(q^2)$ at one loop, which is a smooth function both for time-like and space-like $q^2$.

Numerical integrations, either space-like or time-like, are performed with a \texttt{Mathematica} code employing machine precision numbers and without any symbolic manipulation of the integrand. 
This ensures that we can compare the two cases and use the same code for $\Pi^\lep(q^2)$ as well as for $\Pi^\had(q^2)$.
The numerical values of $\Pi^\had(q^2)$ and the $R$ ratio are provided by the Fortran library \texttt{alphaQED}~\cite{Jegerlehner:2001ca,Jegerlehner:2006ju,Jegerlehner:2011mw}, and \texttt{Rhad}~\cite{Harlander:2002ur} for the regions where perturbative QCD applies, via a mathlink interface.

We make a comparison with the irreducible diagrams calculated with the dispersive method in~\cite{Fael:2019nsf}.
The dispersion relation~\eqref{eqn:dispersionrelationR} effectively replaces the dressed photon propagator with the propagator of a massive gauge boson. These amplitudes are generated by \texttt{FeynArts}~\cite{Hahn:2000kx} with a modified version of the QED model that contains, besides leptons and photon fields, a massive gauge boson with squared mass equal to $z$. 
Later on, the amplitudes are reduced by~\texttt{FormCalc}~\cite{Kuipers:2012rf,Nejad:2013ina} to one-loop tensor coefficients which are calculated by the Fortran library \texttt{Collier}~\cite{Denner:2016kdg} via the \texttt{CollierLink} interface~\cite{Patel:2015tea}.
\texttt{Collier} features dedicated expansions in numerically dangerous regions (small Gram or other kinematical determinants). We particularly benefited from the use of this library because in the numerical evaluation of the dispersive integral~\eqref{eqn:dispersionrelationR} the photon mass $z$  appearing inside the loop can acquire values a few orders of magnitude larger than the typical energy scales of the scattering process.
The numerically stable results provided  by \texttt{Collier} in this treacherous region speeded up the convergence of the dispersive integrals.

We begin by comparing the form factors in eq.~\eqref{eqn:ffintrep}. 
By employing $\Pi^\lep(q^2)$ instead of $\Pi^\had(q^2)$, i.e.\ substituting the hadronic bubble in figure~\ref{fig:HLO} with an electron or a muon loop, we can compare our numerical integration with the analytic results of ref.~\cite{Bonciani:2003ai}, where the QED form factors at two loops were presented. 
In~\cite{Bonciani:2003ai} the vacuum polarization contribution was calculated with the lepton inside the bubble equal to the external one.
\begin{figure}[tb]
  \centering
  \begin{minipage}[t]{\textwidth}
  \centering
    \includegraphics[width=0.45\textwidth]{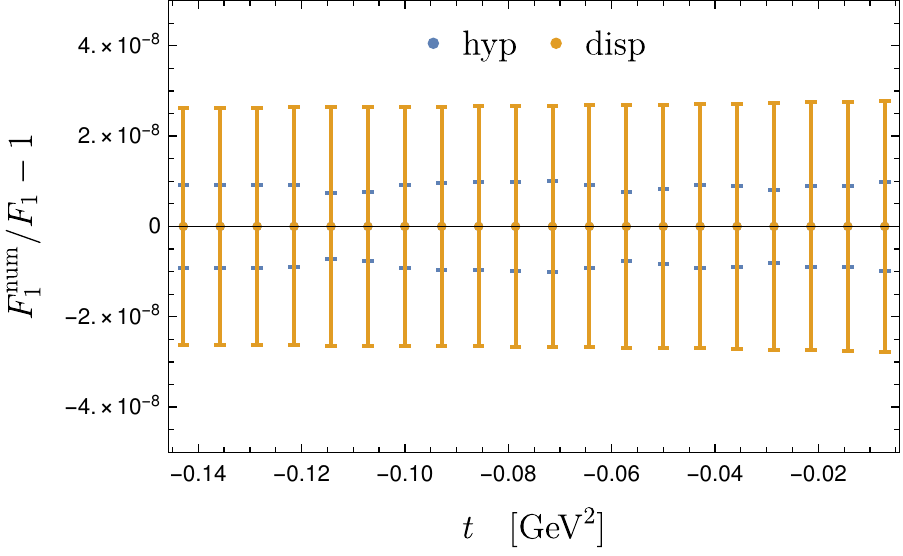}\quad
    \includegraphics[width=0.45\textwidth]{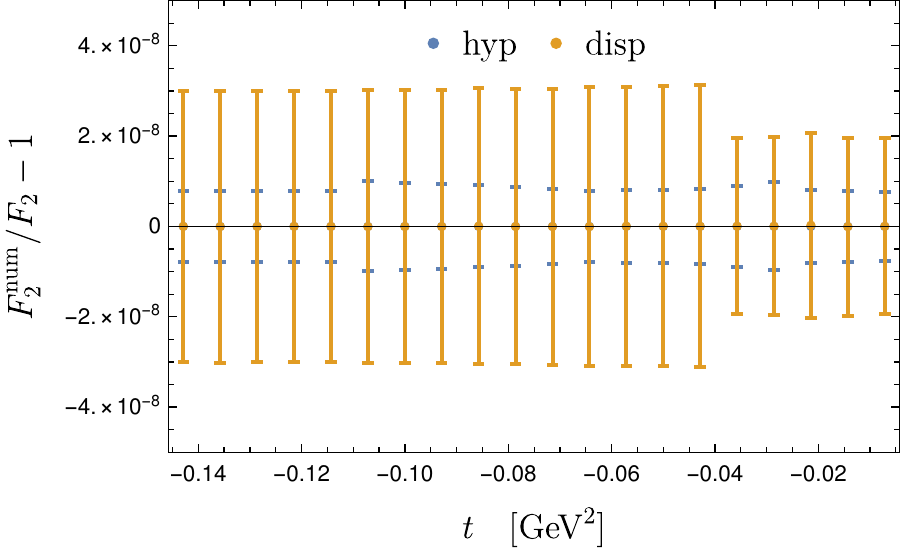}
    \subcaption{Electron vacuum polarization contribution to the electron's form factors.}
    \label{fig:FFlepelec}
  \end{minipage}

  \vspace*{0.05\textwidth}
  \begin{minipage}[t]{\textwidth}
  \centering
    \includegraphics[width=0.45\textwidth]{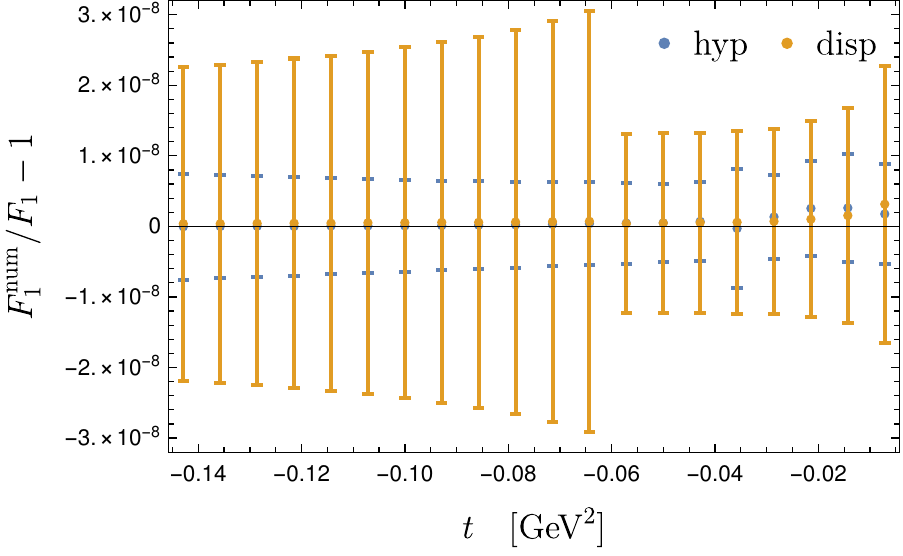}\quad
    \includegraphics[width=0.45\textwidth]{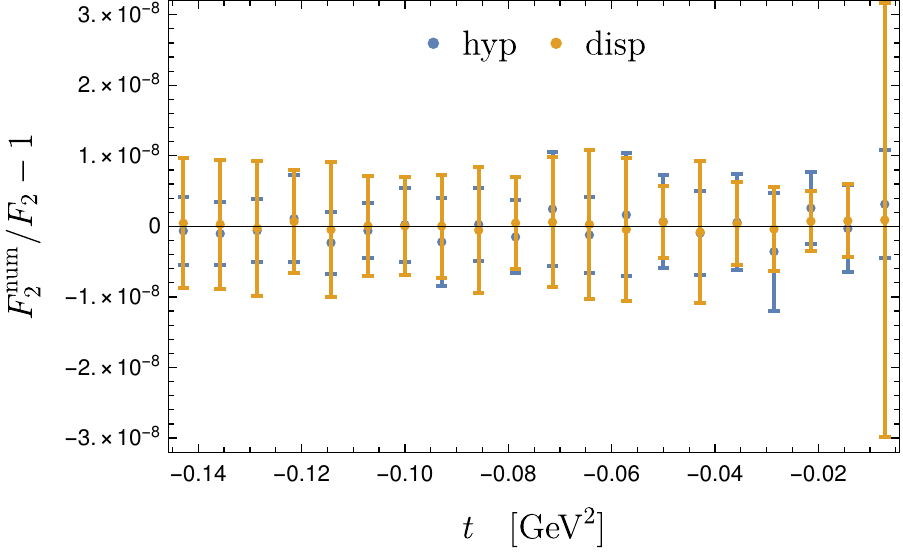}
    \subcaption{Muon vacuum polarization contribution to the muon's form factors.}
    \label{fig:FFlepmuon}
  \end{minipage}
  \caption{Leptonic vacuum polarization contribution to the form factors for $\sqrt{s} = 0.405$~GeV and $-0.142$~Gev$^2 \le t \le 0$~GeV$^2$. 
    Relative difference between the numerical values $F_i^\mathrm{num}$ obtained by the hyperspherical or the dispersive  method and the exact two-loop result $F_i$ in~\cite{Bonciani:2003ai}.
  The error bars show the uncertainty due to numerical integration.}
\end{figure}
The relative difference $(F_i^\mathrm{num}/F_i) -1$ between the form factors calculated numerically with the hyperspherical method, $F_i^\mathrm{num} = F_i^\mathrm{hyp}$, and the exact two-loop result $F_i$ is shown in figure~\ref{fig:FFlepelec} and~\ref{fig:FFlepmuon} for the electron and the muon case, respectively.
The comparison is done for values of the Mandelstam variable $t$ accessible by the \muone experiment at $\sqrt{s}=0.405$ GeV: $-0.142$~GeV$^2 \le t \le 0$~GeV$^2$. 
The error bars are the uncertainty due to the numerical integration. 
Harmonic polylogarithms are evaluated with the \texttt{HPL} package~\cite{Maitre:2005uu,Maitre:2007kp}.
In addition to that, we calculated the QED form factors by employing the dispersion relation~\eqref{eqn:dispersionrelationR} and the analytic expression of $\mathrm{Im} \, \Pi^\lep(q^2)$. The relative difference with $F_i^\mathrm{num}=F_i^\mathrm{disp}$ is shown as well in figure~\ref{fig:FFlepelec} and~\ref{fig:FFlepmuon}. Both methods are in very good agreement with the exact two-loop results, at the level of one part in $10^{-8}$.

\begin{figure}[tb]
  \centering
  \begin{minipage}[t]{\textwidth}
  \centering
    \includegraphics[width=0.45\textwidth]{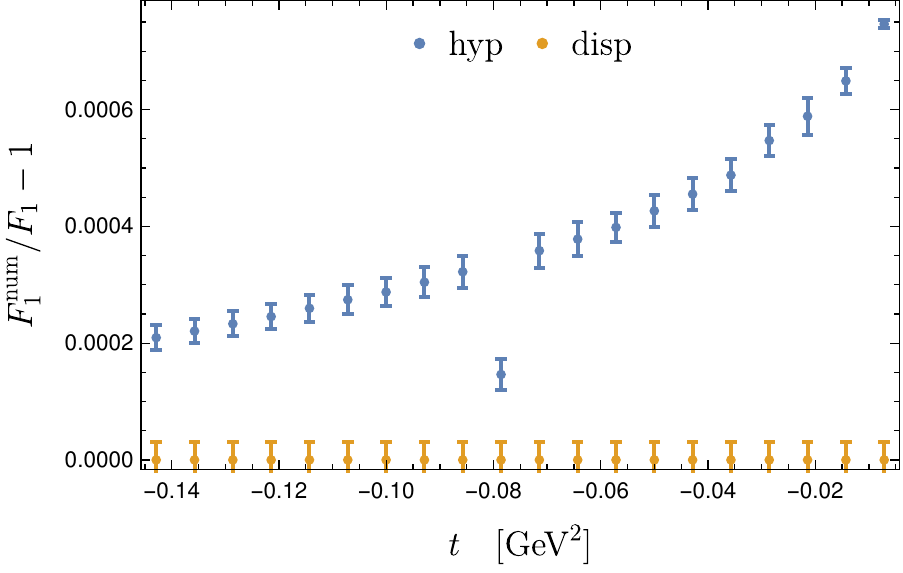}\quad
    \includegraphics[width=0.45\textwidth]{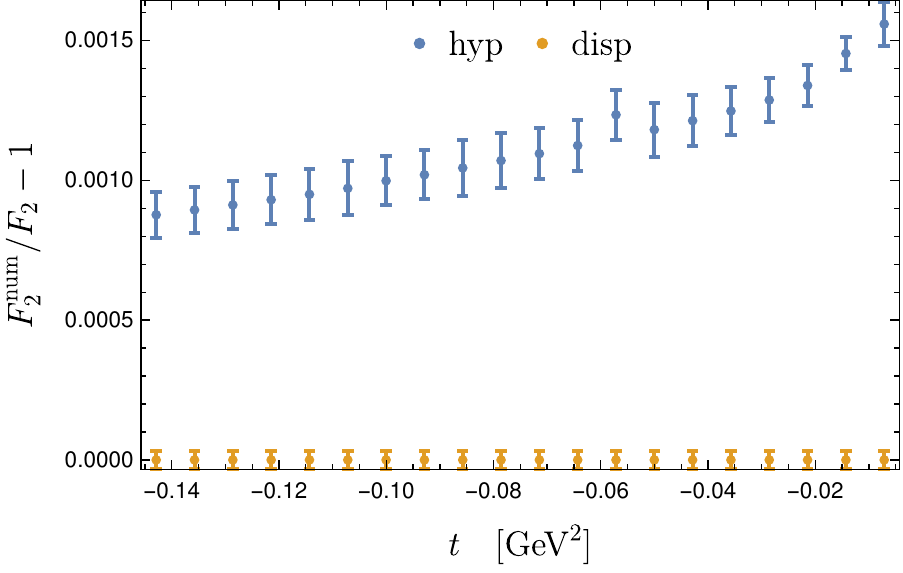}
    \subcaption{Hadronic vacuum polarization contribution to the electron's form factors.}
    \label{fig:FFhadelec}
  \end{minipage}

  \vspace*{0.05\textwidth}
  \begin{minipage}[t]{\textwidth}
  \centering
    \includegraphics[width=0.45\textwidth]{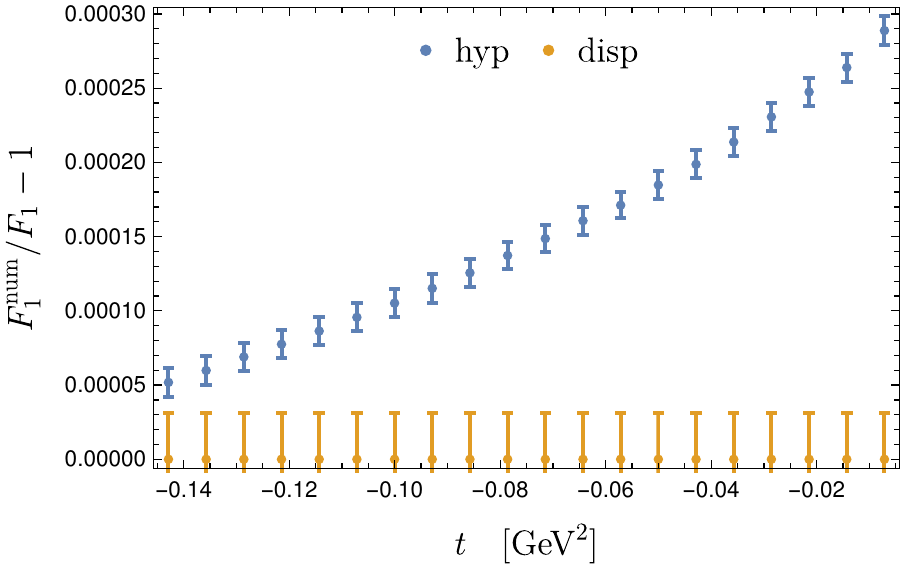}\quad
    \includegraphics[width=0.45\textwidth]{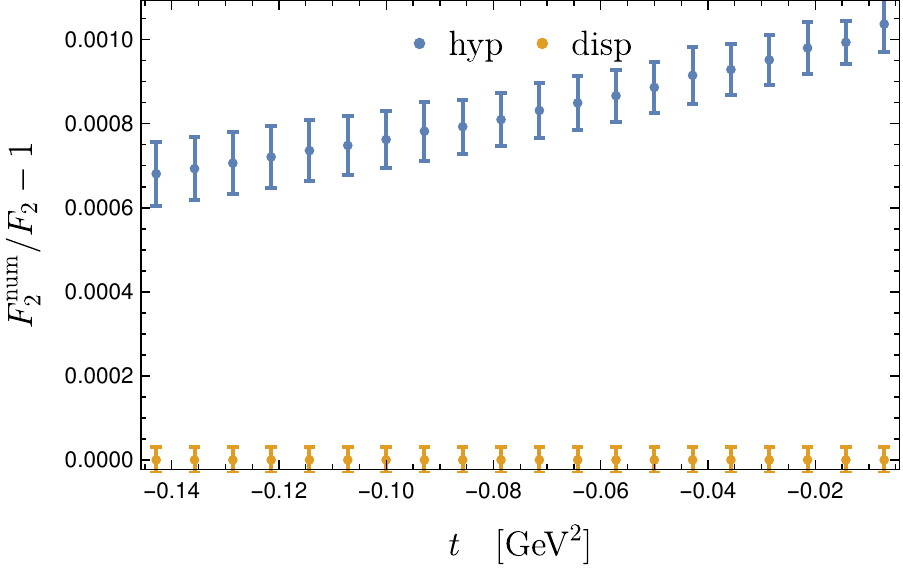}
    \subcaption{Hadronic vacuum polarization contribution to the muon's form factors.}
    \label{fig:FFhadmuon}
  \end{minipage}
  \caption{Hadronic vacuum polarization contribution to the form factors for $\sqrt{s} = 0.405$~GeV and $-0.142$~Gev$^2 \le t \le 0$~GeV$^2$.  
    Relative difference between the values obtained with the hyperspherical and the dispersive  method.
    The result given by the dispersive method is used as normalization factor.
  The error bars show the uncertainty due to the numerical integration.}
  \label{fig:FFhad}
\end{figure}
The comparison with $\Pi^\had(q^2)$ is shown in figure~\ref{fig:FFhadelec} and~\ref{fig:FFhadmuon} for the electron and muon form factors, respectively.
In this case lacking an ``exact'' two-loop expression, we choose as normalization $F_i$ the dispersive method's result.
The values shown in figure~\ref{fig:FFhad} are obtained with the same code employed for $\Pi^\lep(q^2)$, except for the use of $\Pi^\had(q^2)$ and $\mathrm{Im} \, \Pi^\had(q^2)$ instead of the leptonic ones.
We note that with the hadronic vacuum polarization there is a small systematic shift between the numerical values obtained with the two methods, a relative difference of about $10^{-3} - 10^{-4}$. An improvement of the numerical integration error does not change the picture.

\begin{figure}[tb]
  \centering
  \includegraphics[width=0.5\textwidth]{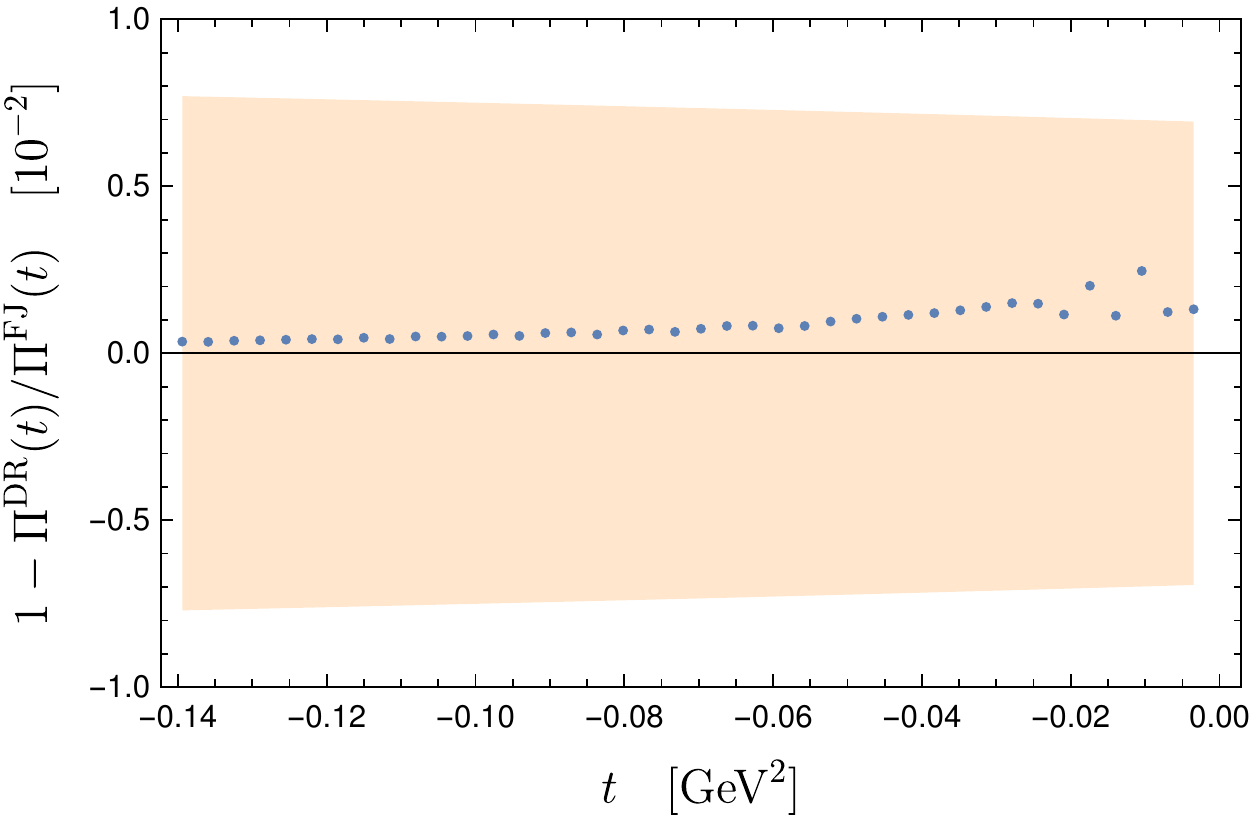}
  \caption{Blue dots are the difference between the hadronic vacuum polarization provided by \texttt{alphaQED}, $\Pi^\mathrm{FJ}(t)$, and the value obtained by direct integration of the dispersion relation with $R(s)$ from the same library, $\Pi^\mathrm{DR}(t)$. The orange band shows the experimental uncertainty on~$\Pi^\mathrm{FJ}(t)$.}
  \label{fig:disprel}
\end{figure}

The source of this shift is the following. 
  The function $\Pi^\had(q^2)$ provided by the library \texttt{alphaQED} is not obtained from a direct integration of the $R$ ratio via~\eqref{eqn:dispersionrelationR}.
It is actually calculated in a different way: first different experiments are integrated separately and then weighted averages of the integrals are taken.   This procedure appears to be more reliable for error estimate, especially in the $\pi \pi$ channel.
The imaginary part provided by \texttt{alphaQED}, i.e.\ the time-like $R$, is obtained by averaging data energy-bin-wise. Therefore the numerical integration of this ``unified'' $R$ can slightly differ from the first procedure since integration and averaging do not commute in general~\cite{privateFred}. 
This effect is shown in figure~\ref{fig:disprel} where we compare, at space-like $t$, the difference between the hadronic vacuum polarization provided by \texttt{alphaQED}, denoted by $\Pi^\mathrm{FJ}(q^2)$, and the values obtained by direct integration of the dispersion relation with $R$ from the same library, denoted by $\Pi^\mathrm{DR}(q^2)$. 
We note a small difference of the order of $10^{-3}$, compatible with the systematic shift appearing in figure~\ref{fig:FFhad}.
The two determinations of $\Pi^\had(q^2)$ are nevertheless in very good agreement within the experimental uncertainty on $\Pi^\mathrm{FJ}(q^2)$, which is also provided by~\texttt{alphaQED} (shown by the orange band in figure~\ref{fig:disprel}).

Also the results of the dispersive and hyperspherical methods are in good agreement taking into account the experimental error from the $R$ ratio. 
The muon and the electron form factor $F_2$ at $t=0$ corresponds to $a_\mu^\mathrm{HLO}$ and $a_e^\mathrm{HLO}$. Their relative uncertainties  are about $0.6\%$~\cite{Jegerlehner:2017zsb,Davier:2017zfy,Keshavarzi:2018mgv}, much larger than the discrepancy appearing in figure~\ref{fig:FFhad}.
One should remind however that the kernel functions employed in the dispersive evaluation of $F_1$ and $F_2$ at $t\neq 0$ are different from $\hat K(s)$ in the \gmt formula~\eqref{eqn:disprel}, so the integration procedure would give in principle a different relative error because the experimental data are weighted differently.
The uncertainty on $a_\mu^\mathrm{HLO}$ and $a_e^\mathrm{HLO}$ ($0.6\%$) must be understood as an order of magnitude of the error at $t\neq 0$ and not as a precise estimate.  
A explicit calculation of the uncertainty for all $t$ is beyond the scope this analysis: it would require the combination of systematic and statistical errors of the data together with their correlation matrices. 
However, even assuming in figure~\ref{fig:FFhad} that the relative error due to $R$ is~$0.1 \%$, which is a factor of six smaller than the uncertainty on $a_\mu^\mathrm{HLO}$ and $a_e^\mathrm{HLO}$, the dispersive and the hyperspherical method would be still in agreement.

\begin{figure}[tb]
  \centering
  \begin{minipage}[t]{0.45\textwidth}
  \centering
    \includegraphics[width=\textwidth]{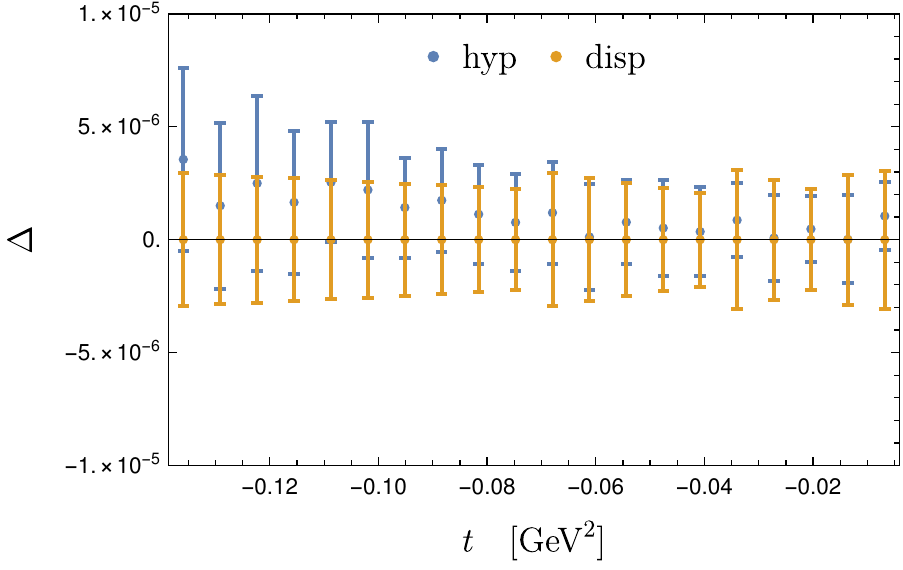}
    \subcaption{Leptonic correction to the uncrossed box.}
    \label{fig:boxuncrossedlep}
  \end{minipage}\quad
  \begin{minipage}[t]{.45\textwidth}
  \centering
    \includegraphics[width=\textwidth]{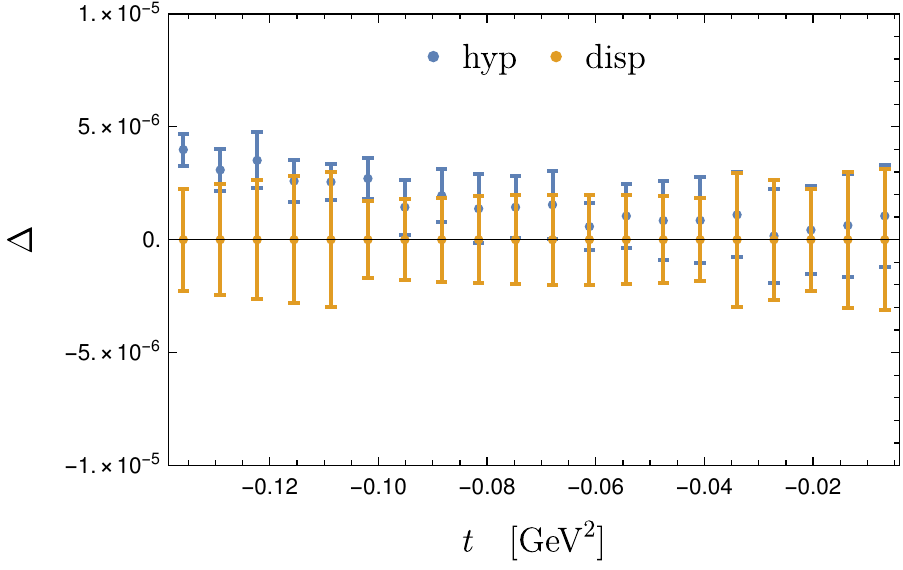}
    \subcaption{Leptonic correction to the crossed box.}
    \label{fig:boxcrossedlep}
  \end{minipage}\\

  \vspace*{0.05\textwidth}
  \begin{minipage}[t]{.45\textwidth}
  \centering
    \includegraphics[width=\textwidth]{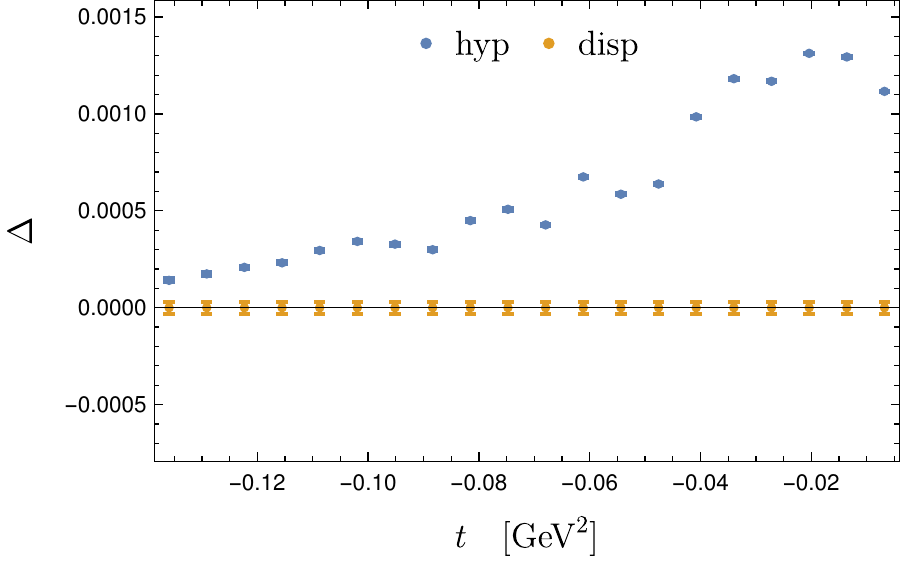}
    \subcaption{Hadronic correction to the uncrossed box.}
    \label{fig:boxuncrossedhad}
  \end{minipage}\quad
  \begin{minipage}[t]{.45\textwidth}
  \centering
    \includegraphics[width=\textwidth]{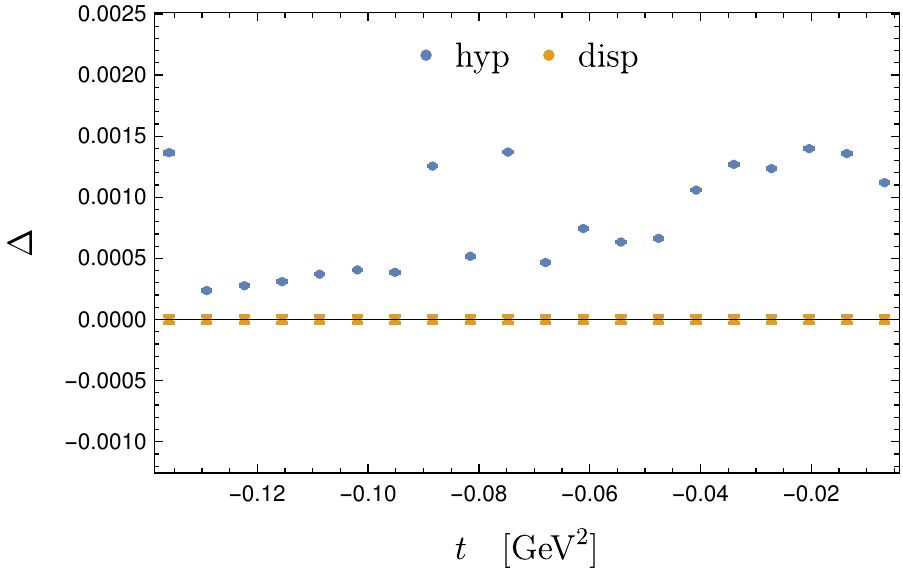}
    \subcaption{Hadronic correction to the crossed box.}
    \label{fig:boxcrossedhad}
  \end{minipage}
  \caption{The relative difference $\Delta$ between the box-Born interferences (with leptonic or hadronic vacuum polarization) calculated by the hyperspherical and the dispersive method for $\sqrt{s} = 0.405$ GeV and $-0.142$~Gev$^2 \le t \le 0$~GeV$^2$.
The result given by the dispersive method is used as normalization factor.
  The error bars show the uncertainty due to the numerical integration.}
  \label{fig:comparisonboxes}
\end{figure}
Let us move on to the box diagrams. 
Figure~\ref{fig:comparisonboxes} shows for $\sqrt{s} = 0.405$~GeV and $-0.142$~Gev$^2 \le t \le 0$~GeV$^2$ the relative difference
\begin{equation}
  \Delta = \frac{|\mathcal{I}^{\mathrm{hyp}} -\mathcal{I}^\mathrm{disp}| }{|\mathcal{I}^\mathrm{disp}|}
\end{equation}
between the Born-virtual interferences $\mathcal{I}=\mathcal{M}_\mathrm{box}\mathcal{M}_\mathrm{Born}^\dagger$ calculated by means of the hyperspherical and the dispersive methods. 
The result given by the dispersive method is chosen as normalization factor in the plots.
The leptonic (hadronic) corrections to the uncrossed and crossed diagram are compared in figure~\ref{fig:boxuncrossedlep} and~\ref{fig:boxcrossedlep} (\ref{fig:boxuncrossedhad} and~\ref{fig:boxcrossedhad}), respectively.
Good agreement is found between the hyperspherical and the dispersive method when the leptonic vacuum polarization function is employed, at the level of one part in $10^{-5}$. The boxes with the hadronic vacuum polarization show also in this case a systematic shift of about~$10^{-3}$ or smaller between the two different calculation, similarly to what we have already observed in the calculation of the form factors. 

For each value of $t$, the dispersive method's result is obtained by performing only one numerical integration: the convolution between the $z$-dependent virtual-Born interference and the imaginary part of the vacuum polarization.
On the contrary, to achieve good numerical stability  with the hyperspherical method we had to evaluate separately for each box topology the $14$ radial integrals $I$. 
Some of the kernel functions are very unstable  around $Q^2 = |t|$ and $Q^2 = +\infty$ and therefore a dedicated series expansion must be employed in these regions.
The IR divergent integrals in eqs.~(\ref{eqn:Iijksub}-\ref{eqn:IDeltaijksub}) with the constant term $\Pi(t)$ in front of it were written in term of one-loop scalar functions and calculated with \texttt{Collier}.

\section{Conclusions \label{sec:conc}}
The present error on the hadronic leading order contribution to the muon \gmt constitutes roughly 50\% of the error budget in the SM prediction.
The \muone experiment proposed at CERN aims at measuring the running of the fine-structure-constant in the space-like region in $\mu$-$e$ scattering and to determine from it $a_\mu^\mathrm{HLO}$ with an error of about $2 \times 10^{-10}$. 
To reach such level of precision it will be necessary to measure the differential cross section with an uncertainty of the order of 10 ppm.
To this end, a Monte Carlo generator with QED and QCD radiative corrections up to NNLO in $\alpha$ must be developed.

In this article we studied the hadronic contributions to the NNLO cross section and we presented a method to evaluate numerically the non-factorizable two-loop diagrams with space-like data for the hadronic vacuum polarization, without making use of the $R$ ratio. In this way the same space-like data measured by \textsc{MUonE}, together perhaps with lattice data and QCD perturbative results, could be exploited to calculate these hadronic corrections. This would allow us to decouple the space-like determination of $a_\mu^\mathrm{HLO}$ from any time-like input.

This work took advantage of the hyperspherical integration method, that was described in section~\ref{sec:hyp}, to express the irreducible vertex and box corrections as a convolution between the vacuum polarization evaluated at negative $q^2$ together with a kernel function obtained by analytic integration of the loop diagrams with respect to the hyperspherical angular variables. 
The vertex corrections were presented in section~\ref{sec:vertex} in terms of QED form factors. In section~\ref{sec:boxes} we showed that each of box contributions can be reduced to a linear combination of 14 integrals which are calculable with the hyperspherical method. Some of these integrals are IR divergent. By making a dedicated subtraction, we managed to remove the IR poles from the integrals explicitly containing the hadronic vacuum polarization and to isolate them in terms that are calculable analytically with standard methods.

Finally, in section~\ref{sec:results} we showed that the numerical evaluation of these irreducible diagrams gives results in agreement with the standard dispersive approach and --- when the analytic expression of $\Pi^\lep(q^2)$ is employed --- in agreement with analytic two-loop vertex results in QED. 
A complete calculation of the hadronic corrections to $\mu$-$e$ scattering at NNLO with the dispersive approach will be presented soon~\cite{Fael:2019nsf}.

\acknowledgments
A special thank to G.\ Colangelo and J.\ Ruiz de Elvira for fruitful discussion about the hyperspherical method, and to T.\ Huber and M.\ Passera for reading the draft and for their valuable suggestions. 
I wish to thank also R.\ Bonciani, C.\ Carloni Calame, F.\ Jegerlehner, S.\ Laporta, P.\ Mastrolia, A.\ Nesterenko, A.\ Primo, E.\ Remiddi, O.\ Tomalak and W.~Torres~Bobadilla for useful discussions and correspondence.
This work was supported by DFG through the Research Unit FOR 1873 ``Quark Flavour Physics and Effective Field Theories'' and by the Mainz Institute for Theoretical Physics (MITP) during the workshop ``The evaluation of the leading hadronic contribution to the muon anomalous magnetic moment''.
Feynman diagrams were drawn with \texttt{Jaxodraw}~\cite{Binosi:2008ig}.

\appendix

\section{One-loop Integral with the Hyperspherical Method: An Example}
In this appendix we present an example of a one-loop calculation with the hyperspherical method and we discuss how to perform the analytic continuation between the Euclidean and the physical region.
We consider, as an example, the loop integral in eq.~\eqref{eqn:d1d3}:
\begin{equation}
  I_{013} = \frac{1}{i \pi^2} 
  \int d^4 q \frac{\Pi^\had(q^2)}{
  (q^2 + i\varepsilon) [(q+p_1)^2-m^2+i \varepsilon] [(q-p_2)^2 -M^2+i \varepsilon]}.
\end{equation}
After continuation of external and internal momenta to the Euclidean region,  Wick rotation and the introduction of hyperspherical coordinates, the loop integral is cast in the following form:
\begin{equation}
  \int d Q^2 \,  Q^2 \, \Pi^\had(-Q^2) \,
  \int
  \frac{d\Omega_Q}{2 \pi^2} 
  \frac{(-1)^3}{
  Q^2 [(Q+P_1)^2+m^2] [(Q-P_2)^2 +M^2]}.
\end{equation}
We expand the propagators as series in Gegenbauer polynomials:
\begin{align}
  \frac{1}{(Q+P_1)^2+m^2} &=
  \frac{Z_1}{|Q ||P_1|}
  \sum_{n=0}^\infty (-Z_1)^n C_n^{(1)} (\hat Q \cdot \hat P_1),
  \label{eqn:sumZ1}\\
  \frac{1}{(Q-P_2)^2+m^2} &=
  \frac{Z_2}{|Q|| P_2|}
  \sum_{n=0}^\infty Z_2^n C_n^{(1)} (\hat Q \cdot \hat P_2),
  \label{eqn:sumZ2}
\end{align}
where 
\begin{align}
  Z_1 &= \frac{Q^2+P_1^2+m^2-\lambda^{1/2}(Q^2,P_1^2,-m^2)}{2 |Q|| P_1|},
  \label{eqn:defZ1}\\
  Z_2 &= \frac{Q^2+P_2^2+M^2-\lambda^{1/2}(Q^2,P_2^2,-M^2)}{2 |Q ||P_2|}.
  \label{eqn:defZ2}
\end{align}
We perform the angular integration by making use of the orthogonality property~\eqref{eqn:orthogonalitycondition}:
\begin{equation}
  \int
  \frac{d\Omega_Q}{2 \pi^2} 
  \frac{1}{
  [(Q+P_1)^2+m^2] [(Q-P_2)^2 +M^2]}= 
  \frac{-1}{Q^2 |P_1|| P_2|}
  \sum_{n=0}^\infty
  \frac{(-Z_1 Z_2)^{n+1}}{n+1}C_n^{(1)} ( \hat P_1 \cdot \hat P_2).
\end{equation}
The series in the expression above can be calculated by defining $z=(-Z_1 Z_2)$ and by taking the derivative w.r.t.\ $z$, that yields: 
\begin{equation}
  \frac{d}{dz} \sum_{n=0}^\infty \frac{z^{n+1}}{n+1}C_n^{(1)} ( \hat P_1 \cdot \hat P_2)
  =
  \sum_{n=0}^\infty z^n  C_n^{(1)} ( \hat P_1 \cdot \hat P_2) =
  \frac{1}{1-2 \tau z + z^2},
\end{equation}
where $\tau = \hat P_1 \cdot \hat P_2$. We then take the primitive and we impose the boundary condition $ \sum_n \frac{z^{n+1}}{n+1} C_n^{(1)} = 0$ at $z=0$.  So the series is:
\begin{align}
  \sum_{n=0}^\infty
  \frac{z^{n+1}}{n+1}C_n^{(1)} ( \hat P_1 \cdot \hat P_2) &=
  \frac{1}{ \sqrt{1-\tau^2}} 
  \left[
  \arctan \left( \frac{z-\tau}{\sqrt{1-\tau^2}} \right)
  -\arctan \left( \frac{-\tau}{\sqrt{1-\tau^2}} \right)
  \right] \notag \\
  & = 
  \frac{1}{ \sqrt{1-\tau^2}} 
  \arctan \left( \frac{z\sqrt{1-\tau^2} }{1-z\tau} \right) \, ,
  \label{eqn:step1}
\end{align}
where we used the addition formula $\arctan(x)-\arctan(y) = \arctan(\frac{x-y}{1+xy} )$.

Having performed the angular integrations, the loop integral takes the form:
\begin{equation}
  I_{013} =
  - \int_0^{+\infty} dQ^2  \,
    \Pi^\had(-Q^2) \,
    f(Q^2,P_1^2,P_2^2,\tau) \, .
    \label{eqn:I013v2}
\end{equation}
Since ultimately we are interested in the answer for time-like $P_1^2$ and $P_2^2$ we have to perform the analytic continuation before the $Q^2$-integration.
The most important point one has to check is whether any singularity crosses the integration path in the $Q^2$ complex plane when $P_1^2$ and $P_2^2$ are continued to negative values. %

\begin{figure}[thb]
  \centering
  \includegraphics[width=0.7\textwidth]{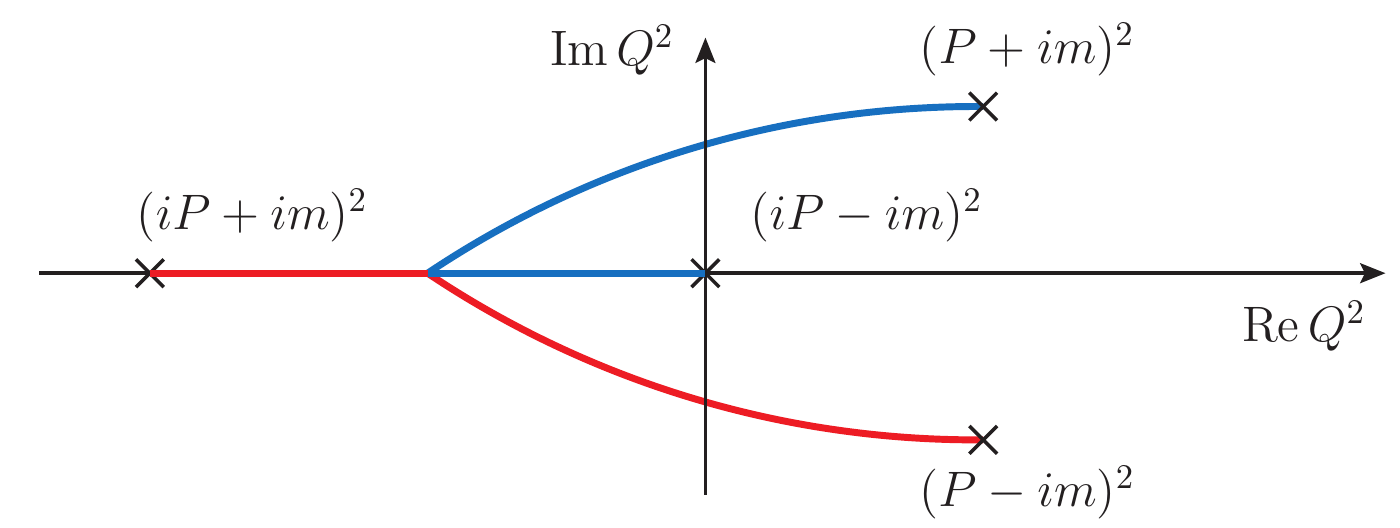}
  \caption{Location of the branch points of $Z_1$, in the $Q^2$ plane. The path shows how these branch points moves as $P_1^2$ is varied from a positive value to $-m^2$.}
  \label{fig:figapp}
\end{figure}

Barring the poles coming from the divergences in the infinite sums in~\eqref{eqn:sumZ1} and~\eqref{eqn:sumZ2}, which does not affect this analysis, the integrand is meromorphic in the variables $Q^2$, $P_1^2$ and $P_2^2$ except for the square roots in the $Z$ variables~\eqref{eqn:defZ1} and~\eqref{eqn:defZ2}.
Let's now study the behaviour of $Z_{1,2}$ when $P_{1,2}^2$ is continued from positive quantities to negative on-shell values $P_1^2=-m^2$ and $P_2^2=-M^2$. 
In the $Q^2$ complex plane, $Z_1$ ($Z_2$) has branch points at $Q^2 = (P_1 \pm i m)^2$ ($Q^2 = (P_2 \pm i M)^2$). 
At the beginning $P_1$ and $P_2$ are real and positive and the integration is performed from $0$ up to $\infty$. 
Figure~\ref{fig:figapp} shows the path of the branch points as $P_{1}^2$ is varied to $-m^2$\footnote{In the original continuation to the Euclidean region we let the energy $p_0$ to acquire a phase $e^{i\phi}$, which is then varied from 0 to $\pi/2$. Therefore $P^2$ moves from positive values to the negative ones passing,  in the $P^2$ complex plane, below the real axis.}. 
When $P_1^2=0$ they are located at $Q^2 = -m^2$. When $P^2_1$ is continued to negative values, one of the branch point moves to the left and the other one reaches the origin at $P^2_1=-m^2$. The branch points of $Z_2$ behave in the same way.
None of the singularities crosses the integration path and therefore we can continue $P_1^2$ ($P_2^2$) to $-m^2$ ($-M^2$) without distorting the $Q^2$ contour.
Note however, if we had to continue $P_1^2$ to a value larger than $m^2$, one of the branch point would have crossed the positive real axis and  we would have needed to distort the contour to get the correct continuation of the integral (see also the discussion in ref.~\cite{Roskies:1990ki}).

The Euclidean result of the angular integration can be then continued to the on-shell configuration by setting $P_1^2 = -m^2$, $P_2^2 = -M^2$ and $(P_1+P_2)^2 =  P_1^2+P_2^2 + 2 |P_1|| P_2| \tau = -s$,  keeping in a first step $(M-m)^2 < s < (M+m)^2$ in order to leave the square roots in~\eqref{eqn:step1} real valued. 
In a second step, we continue $s$ to the physical region $s + i \varepsilon > (M+m)^2$, giving to it a small (positive) imaginary part. 
As in~\eqref{eqn:step1} the square root $\sqrt{1-\tau^2}$ becomes $i \sqrt{\tau^2-1}$, we can rewrite the arctangent in terms of $L(z)$ (the hyperbolic inverse tangent) via the identity: $\arctan (iz) = i L(z)$. Eventually the kernel function appearing in the integral~\eqref{eqn:I013v2} can be cast in the following form:
\begin{multline}
  f(Q^2,-m^2,-M^2,s) = 
  \frac{-2}{Q^2 \lambda^{1/2}(s,M^2,m^2)}  \\[5pt]
  \times 
  L\left( 
  \frac{\lambda^{1/2}(s,M^2,m^2)} 
  {s-M^2-m^2 - 8 M^2 m^2 \Big/ 
  \left[ Q^2
  \left( 1-\sqrt{1+\frac{4m^2}{Q^2}} \right)
\left( 1-\sqrt{1+\frac{4M^2}{Q^2}} \right)\right]}
  - i \varepsilon
  \right) \, .
  \label{eqn:d1d3v2}
\end{multline}
Let's analyze this formula. As expected, the function $L$ has an imaginary part for $s> (M+m)^2$ since $s$ is continued above the physical threshold. 
For real $z$, the function $L(z)$ acquires an imaginary part when $|z|>1$, which happens in the bounded region $0< Q^2 < \lambda(s,M^2,m^2)/s$. 
Eq.~\eqref{eqn:d1d3v2} provides also the result for $s<(M-m)^2$, which corresponds to a $u$-channel configuration with $s$ substituted by $u$. In this case it gives the formula~\eqref{eqn:d1d4} for the crossed box. 
One can verify that no imaginary part is developed if $s = u < (M-m)^2$, since $s$ is below the physical threshold.
Indeed the argument of  $L$ is monotonically increasing for $Q^2 \to +\infty$ and it is bounded between $-\sqrt{(s-(M-m)^2)/(s-(M+m)^2)}$ (at $Q^2=0$) and zero (at $Q^2 \to + \infty$). 

Finally one can show that eq.~\eqref{eqn:d1d3v2} is equivalent to the expression in~\eqref{eqn:d1d3} and~\eqref{eqn:d1d4}, that were derived from eq.~$10$ in ref.~\cite{Laporta:1994mb}, by making use of the identities $L(u)+L(w) = L(\frac{u+w}{1+u w})$ and $L(z) = L(1/z) + i\pi/2$ (for $|z|>1$). The formula~\eqref{eqn:d1d3v2} in the equal mass case, i.e.\ $m^2=M^2$, appears also in the calculation of the vertex form factors in eq.~\eqref{eqn:f1} and~\eqref{eqn:f2}.

\label{Bibliography}
\bibliographystyle{JHEP}
\bibliography{BIB}
\end{document}